\documentclass[]{aa}  

\usepackage{graphicx}
\usepackage{natbib}
\bibpunct{(}{)}{;}{a}{}{,}
\usepackage{txfonts}
\usepackage{supertabular}
\usepackage{ulem}
\usepackage{longtable,lscape}

\begin{document} 

  \title{HD 51844: An Am $\delta$ Scuti in a binary showing periastron brightening\thanks{Based on observations obtained with the HERMES spectrograph attached to the Mercator Telescope  which is operated on the island of La Palma by the University of Leuven (IvS) at the Spanish Observatorio del Roque de los Muchachos of the Instituto de Astrof\'isica de Canarias. The HERMES spectrograph is supported by the Fund for Scientific Research of Flanders (FWO), Belgium , the Research Council of K.U.Leuven, Belgium, the Fonds National de la Recherche Scientifique (FNRS), Belgium, the Royal Observatory of Belgium, the Observatoire de Gen\`eve, Switzerland and the Th{\"u}ringer Landessternwarte Tautenburg, Germany.
  Based on CoRoT space-based photometric data; the CoRoT space mission was developed and operated by the French space agency CNES, with the participation of ESA’s RSSD and Science Programmes, Austria, Belgium, Brazil, Germany, and Spain. Based on observations collected at La Silla Observatory, ESO (Chile) with the HARPS spectrograph at the 3.6-m telescope, under programme LP185.D-0056.  } }

   \titlerunning{\object{HD\,51844}: $\delta$ Scuti in an eccentric binary system}

   \author{M. Hareter
          \inst{1}
          \and M. Papar\'o\inst{1}
          \and W. Weiss\inst{2} 
          \and A. Garc{\'i}a Hern{\'a}ndez\inst{3}           
          \and T. Borkovits\inst{4}$^,$\inst{5}
          \and P. Lampens\inst{6}
	  \and M. Rainer\inst{7}
          \and P. De Cat\inst{6}
          \and P. Marcos-Arenal\inst{8}
          \and J. Vos\inst{8}
          \and E. Poretti\inst{7}
          \and A. Baglin\inst{9}
          \and E. Michel\inst{9}
          \and F. Baudin\inst{10}
          \and C. Catala\inst{9}
          }

   \institute{Konkoly Observatory, MTA CSFK, Konkoly Thege M. \'ut 15-17., H-1121 Budapest, \\
              \email{hareter@konkoly.hu}
         \and Institut f\"ur Astrophysik, Universit\"at Wien, T\"urkenschanzstra\ss e 17, 1180 Wien, Austria
         \and Centro de Astrof{\'i}sica, Universidade do Porto, Rua das Estrelas, 4150-762 Porto, Portugal 
         \and Baja Astronomical Observatory, H-6500 Baja, Szegedi \'ut, Kt. 766, Hungary 
	 \and ELTE Gothard-Lend\"ulet Research Group, H-9700 Szombathely, Szent Imre herceg \'ut 112, Hungary
         \and Koninklijke Sterrenwacht van Belgi\"e, Ringlaan 3, 1180 Brussel, Belgium 
         \and INAF Osservatorio Astronomico di Brera, Via E. Bianchi 46, 23807 Merate (LC),Italy 
         \and Instituut voor Sterrenkunde, KU Leuven, Celestijnenlaan 200D, 3001 Leuven, Belgium
         \and LESIA, Observatoire de Paris, site de Meudon, UMR CNRS 8109, associ{\'e} {\`a} l'Universit{\'e} Pierre et Marie Curie (Paris 6) et {\`a} l'Universit{\'e} Paris-Diderot (Paris 7).  5 place J. Janssen, 92195 Meudon Cedex, France 
         \and Institut d'Astrophysique Spatiale, UMR8617, CNRS, Universit\'{e} Paris XI, B\^{a}timent 121, 91405 Orsay Cedex, France
}
   \date{}

 
  \abstract
 {Pulsating stars in binary systems are ideal laboratories to test stellar evolution and pulsation theory, since a direct, model-independent determination of component masses is possible. 
The high-precision CoRoT photometry allows a detailed view of the frequency content of pulsating stars, enabling detection of patterns in their distribution. The object \object{HD\,51844}\ is such a case showing periastron brightening instead of eclipses.}
{We present a comprehensive study of the \object{HD\,51844}\ system, where we derive physical parameters of both components, the pulsation content and frequency patterns. Additionally, we obtain the orbital elements, including masses, and the chemical composition of the stars.}
{Time series analysis using standard tools was employed to extract the pulsation frequencies.   
Photospheric abundances of 21 chemical elements were derived by means of spectrum synthesis. We derived  orbital elements both by fitting the observed radial velocities and the light curves, and we did asteroseismic modelling as well.}
{We found that \object{HD\,51844}\ is a double lined spectroscopic binary. 
The determined abundances are consistent with $\delta$ Delphini classification. We determined the orbital period (33.498$\pm$0.002\,d), the eccentricity 
(0.484$\pm$0.020), the mass ratio (0.988$\pm0.02$), and the masses to 2.0$\pm$0.2\,$M_{\sun}$ for both components.
Only one component showed pulsation. Two p modes ($f_{22}$ and $f_{36}$) and one g mode ($f_{orb}$) may be tidally excited. Among the 115 frequencies, we detected triplets due to the frequency modulation, frequency differences connected to the orbital period, and unexpected resonances (3:2, 3:5, and 3:4), which is a new discovery for a $\delta$ Sct\ star. The observed frequency differences among the dominant modes suggest a large separation of 2.0 -- 2.2\,d$^{-1}$, which are consistent with models of mean density of 0.063\,g\,cm$^{-3}$, and with the binary solution and TAMS evolutionary phase for the pulsating component. The binary evolution is in an intermediate evolutionary phase; the stellar rotation is super-synchronised, but circularisation of the orbit is not reached.}
{}

   \keywords{stars:pulsation --
                stars:$\delta$ Sct\ stars --
                stars: binaries --
                techniques:photometry --
                techniques:spectroscopy
               }

   \maketitle
%

\section{Introduction}

Pulsating components of binary stars are most valuable targets to test various theories of stellar modelling, provided that the component properties (e.g., masses, radii) can be determined with sufficient accuracy. The characteristic behaviour of binary systems may also be investigated with the help of pulsating components of binary systems. Even though eclipsing binaries showing additional periastron brightening have been discussed for a long time \citep{Roberts06}, it is space photometry, which is a method sensitive enough to show periastron brightening effects of non-eclipsing binaries. The latter objects, which show even more complicated light variations than those present in eclipsing stars were found in {\it Kepler} data and reported by \citet{Thompson12}. However, similar cases have also been found from ground-based observations \citep{Fekel11}. Light curves indicating periastron brightening effects of non-eclipsing binaries can be explained by eccentric orbit resulting in tidal deformation of the stars and/or reflected light.

The object \object{HD\,51844}\ shows the periodic brightening event of a heartbeat star. The phase diagram of \object{HD\,51844}\ resembles that of KIC 4847343 or KIC 8264510 shown in \citet{Thompson12}; hence, \object{HD\,51844}\ is the first recognised case of a heartbeat star among the CoRoT targets. Tidally induced pulsation in g modes with integer multiples of the orbit frequency is frequently observed among such stars of which KOI-54 represents an instructive example for such stars \citep{Welsh11}.

The object \object{HD\,51844}\ (= CoRoT 1043,V = 8.59) is an ideal test-case for various theories. It was observed in the CoRoT LRa02 (from Nov. 11, 2008 to March 11, 2009) for 117 days of uninterrupted time-base as one of ten Seismo targets. Additionally, the star was observed in the Large Program LP 185.D with ESO using the HARPS instrument mounted on the 3.6-m telescope at La Silla. During these observations, \object{HD\,51844}\ was discovered to be a binary \citep{Hareter14}. Furthermore, 19 HERMES spectra were obtained from March to April 2013 covering a large fraction of the binary orbit phase. Five groups of two consecutive McDonald spectra were also obtained in late March 2013.  

The CoRoT light curve reveals multi-mode pulsation typical for $\delta$ Sct\ stars.
Such stars are pulsating intermediate mass stars (1.5 to 3.0 $M_{\sun}$) on or near the main sequence. They pulsate in radial and non-radial low-order p modes, with frequencies ranging typically from roughly 5 to 70\,d$^{-1}$. A detailed review on $\delta$ Sct\ stars can be found in \citet{Breger00, Aerts10}.

The star is listed in the General Catalogue of Ap and Am stars \citep{Renson09} and was originally reported as CrII-EuII by \citet{Muench52}, who used the MKK stellar classification system of \citet{Morgan43}. It is also listed as a Fm $\delta$ Delphini type by \citet{Houk99} in the Michigan catalogue of two-dimensional spectral types for the HD Stars, vol. 5. The $\delta$ Del stars are characterised by a similar chemical abundance pattern as the Am stars but without their typical underabundances of Sc and Ca. A large fraction of Am stars are components of  binary systems \citep{Carquillat07}. Unfortunately, no HIPPARCOS parallax is available.
 
In Sec.\,\ref{sec-lightcurve}, we present the frequency analysis of the CoRoT Seismo photometry, including a detailed study of the regularities of the $\delta$ Sct\ pulsation. In Sec.\,\ref{sec-spec}, we describe the analysis of the spectroscopy, including the abundance analysis and the radial velocity measurements. In Sec.\,\ref{sec-orbit} we derive the orbital solution of the binary and the determination of the stellar parameters. Section\,\ref{sec-model} presents the seismic modelling of the pulsator. Section\,\ref{sec-discussion} discusses the regularity of the p modes, the discrepancies between the stellar parameters obtained from spectroscopy, and orbital and seismic modelling. The evolutionary status of the components and the binary orbit are also discussed. Finally, a summary is given in Sec.\,\ref{sec-conc}.

\section{$\delta$ Sct\ pulsation}
\label{sec-lightcurve}
\subsection{Treatment of the raw data}\label{ssec-freqanal-data}

\begin{figure}
   \centering
   \includegraphics[width=\hsize]{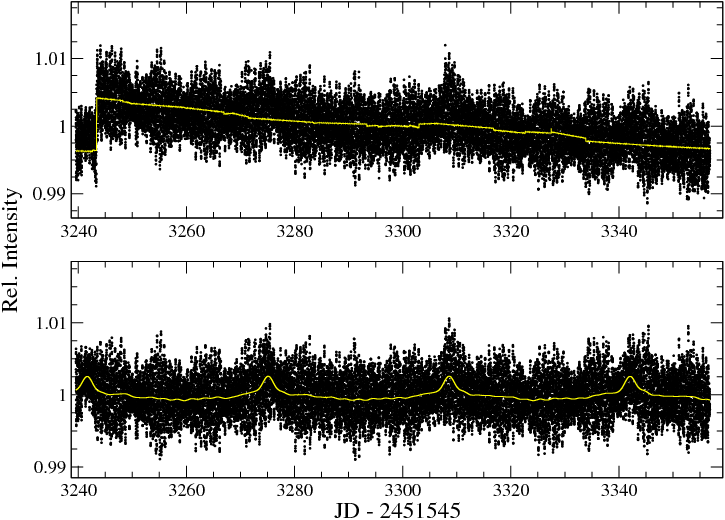}
      \caption{CoRoT N2 light curve of \object{HD\,51844}\ (top panel), the inserted curve shows the final trend curve which was subtracted. The resulting light curve is shown in the bottom panel with the binary light curve indicating the periodic brightening event.}
      \label{fig-hd5-lc}
\end{figure}

We used the N2 data of the CoRoT Seismo light curve for our investigation. We excluded data points, which were flagged as interpolated, but kept all other data points. The jump and outliers were removed in an iterative approach, where the first step was to subtract a linear regression on the whole data set. The next step involved prewhitening with the $\delta$ Sct\ frequencies; in this case, all significant frequencies from 5 to 25 d$^{-1}$. As a result, a clear periodic brightening event became evident. Its period was preliminarily determined using the Phase Dispersion Minimisation method \citep{Stellingwerf78}. This periodic brightening event was also removed by subtracting a multi-sine fit using $f_{orb}$ and 15 harmonics. This latter procedure left the residuals only with the noise and artefacts.

We rectified these residuals in turn by fitting piecewise linear regressions and performing a 3$\sigma$ outlier clipping. We removed the jump at the beginning by adding the appropriate difference to the average residuals (i.e., 0.0018 in relative intensity) and removed the trend curve (unrectified residuals minus rectified residuals) from the N2 data. The top panel in Fig\,\ref{fig-hd5-lc} shows the untreated N2 data and the subtracted trend curve. We also applied the sampling of the residuals, which is reduced by the clipping of outliers, to the N2 data. The resulting cleaned light curve is shown in the bottom panel of Fig\,\ref{fig-hd5-lc}. The overplotted line shows the periodic brightening event.

The $\delta$ Sct\ pulsation and the binary light variation are well separated in the Fourier space, which allows disentangling the light variation due to the binary effect and that due to the pulsation in the time domain. To investigate the pulsation content with high accuracy, we constructed a preliminary binary light curve by a multi-sine fit using the binary orbit frequency and 15 harmonics. We subtracted the latter from the cleaned light curve, resulting in a pure pulsation light curve, which is investigated in this section in detail.

\subsection{Frequency analysis of the pulsational signal}\label{ssec-freqanal-signal}

\begin{table}[h!]
\caption{The 55 most significant independent frequencies of \object{HD\,51844}\ to the A $>$ 0.05 mmag. The numbers between brackets give the first uncertain digit of the Fourier parameters. The full list of 115 significant frequencies can be found in the online material.}
\label{tabl-freq}
\centering
\begin{tabular}{lrcrrrrrrrrrr}
\hline\hline
No. & \multicolumn{1}{c}{Frequency} & \multicolumn{1}{c}{Ampl.} & \multicolumn{1}{c}{Phase} &
\multicolumn{1}{c}{S/N} \\ 
& \multicolumn{1}{c}{(d$^{-1}$)} & \multicolumn{1}{c}{(mmag)} & \multicolumn{1}{c}{(rad)} & \\
\hline
1 & 12.21284(4) & 2.26(2) & 0.1683(8) & 419.99 \\
2 & 7.05419(8) &   2.14(2) & 0.2278(9) & 431.56  \\
3 & 6.94259(6) & 1.83(2) & 0.5071(6) & 371.62 \\
4 & 8.14077(3) & 1.47(4) & 0.150(7) & 282.77 \\
5 & 6.75568(5) & 0.61(5) & 0.797(7) & 125.20 \\
6 & 11.95900(1) & 0.58(4) & 0.388(8) & 108.89 \\
7 & 11.56613(9) & 0.53(7) & 0.512(1) & 95.27 \\
8 & 10.3695(3) & 0.37(1) & 0.476(9) & 66.53 \\
9 & 11.3373(2) & 0.36(5) & 0.440(8) & 65.74 \\
10 & 12.0898(2) & 0.35(7) & 0.554(2) & 65.86 \\
11 & 10.3528(3) & 0.32(3) & 0.292(2) & 57.71 \\
12 & 10.7233(6) & 0.30(6) & 0.970(0) & 55.41 \\
13 & 10.8483(5) & 0.27(7) & 0.957(2) & 49.87 \\
14 & 10.6819(2) & 0.26(1) & 0.687(6) & 47.53 \\
15 & 7.4543(5) & 0.23(7) & 0.359(2) & 46.27 \\
16 & 11.7759(5) & 0.22(9) & 0.230(9) & 40.85 \\
17 & 12.5201(9) & 0.19(1) & 0.506(8) & 36.23 \\
18 & 11.4296(6) & 0.18(3) & 0.132(8) & 32.48  \\
19 & 6.5868(9) & 0.17(2) & 0.360(1) & 35.92 \\
20 & 12.7545(3) & 0.17(2) & 0.064(6)& 33.45 \\
21 & 11.8311(5) & 0.16(2) & 0.07(6) & 29.07 \\
22 & 9.9707(4) & 0.16(3) & 0.86(1) & 28.81 \\
23 & 7.2190(1) & 0.15(5) & 0.94(3) & 31.35 \\
24 & 9.7186(2) & 0.14(8) & 0.02(8) & 27.23 \\
25 & 11.2958(9) & 0.14(3) & 0.90(8) & 25.63 \\
26 & 14.8846(6) & 0.13(9) & 0.70(6) & 24.64 \\
27 & 9.3199(5) & 0.12(9) & 0.85(0) & 23.70 \\
28 & 8.8682(8)  & 0.13(8) & 0.07(9) & 26.47 \\
29 & 8.8818(8) & 0.12(2) & 0.17(3) & 23.42 \\
30 & 12.1572(4) & 0.11(2)  & 0.25(6) & 20.65 \\
31 & 12.2856(7) & 0.10(9)  & 0.09(6) & 20.20 \\
32 & 10.0247(7) & 0.10(1)  & 0.25(3) & 17.91 \\
33 & 8.3888(4) & 0.09(9) & 0.45(5) & 19.06 \\
34 & 12.4223(7) & 0.10(0)  & 0.69(9) & 18.72 \\
35 & 7.3205(4) & 0.09(2)  & 0.26(9) & 18.32 \\
36 & 9.5227(3) & 0.08(6)  & 0.73(0) & 15.67 \\
37 & 10.6712(7) & 0.08(6) & 0.72(6) & 15.71 \\
38 & 10.5005(0) & 0.08(3)  & 0.06(2) & 14.96 \\
39 & 7.9010(9) & 0.08(4)  & 0.82(3) & 16.07 \\
40 & 11.0848(4) & 0.07(6)  & 0.28(5) & 13.22 \\
41 & 11.2855(7) & 0.08(0)  & 0.60(2) & 14.28 \\
42 & 12.2430(9) & 0.07(2)  & 0.58(3) & 13.33 \\ 
43 & 7.5712(2) & 0.07(1)  & 0.60(1) & 13.86 \\
44 & 12.1830(0) & 0.07(1)  & 0.96(0) & 13.09 \\
45 & 7.6520(5) & 0.06(7)  & 0.83(3) & 12.97 \\
46 & 6.7898(5) & 0.06(3)  & 0.04(9) & 12.85 \\
47 & 8.2485(4) & 0.06(7)  & 0.18(1) & 12.75 \\
48 & 9.0977(1) & 0.06(3)  & 0.90(0) & 11.82 \\
49 & 12.7807(9) & 0.06(0) & 0.55(3) & 11.63 \\
50 & 12.3208(9) & 0.05(9) & 0.53(2) & 11.00 \\
51 & 10.8197(9) & 0.05(5) & 0.12(1) & 9.83 \\
52 & 6.9123(6) & 0.05(7) & 0.59(7) & 11.52 \\ 
53 & 7.0245(0) & 0.05(3) & 0.53(5) & 10.66 \\
54 & 8.1106(2) & 0.05(2) & 0.96(6) & 10.02 \\
55 & 8.0741(5) & 0.05(1) & 0.02(1) & 9.85 \\

\hline
\end{tabular}
\end{table}

\begin{figure}
 \centering
\includegraphics[angle=00,width=9cm]{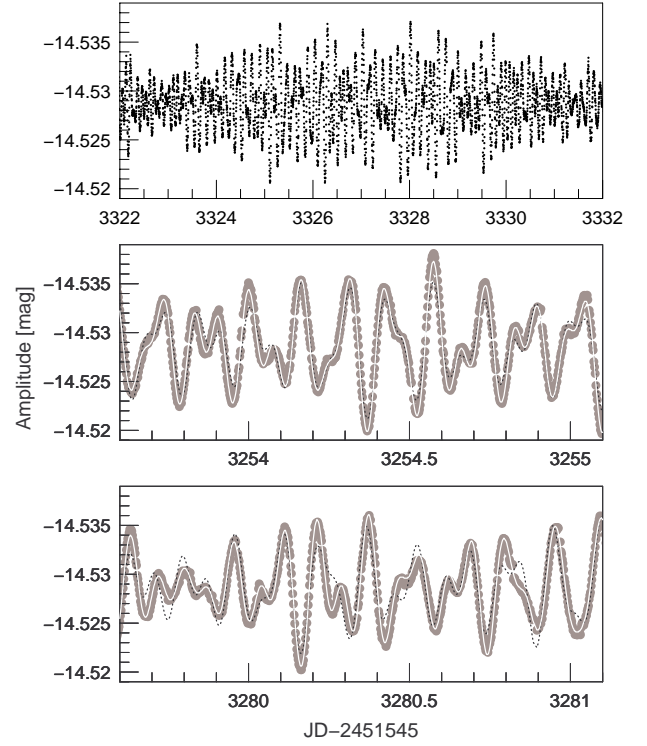}
 \caption{Typical features in the light curve of \object{HD\,51844}. The full amplitude of the cycles is in a 0.015 mag range. The ten-day long beat is caused by two/three dominant modes. Fits with four (black dashed line) and 115 (white continuous line) frequencies are given in the zoomed panels. Observational points are given by large size grey filled circles, serving as a background for the  fit with 115 frequencies.}
\label{fig-light}
\end{figure}

We applied a standard frequency analysis to the pure pulsation light curve to obtain the final pulsational frequencies. Although this Seismo target was sampled in every 32 s and the raw dataset contained
302 107 points, we carried out the analysis on the dataset binned in 3 minutes (55 201 points). Typical segments of the light curve, a ten-day long beating feature, abrupt changes in the amplitude, and the
curvature of the consecutive cycles are shown in Fig.\,\ref{fig-light}. The peak-to-peak amplitude is of the order of 0.015 mag.  

\begin{figure}
 \centering
   \includegraphics[width=\hsize]{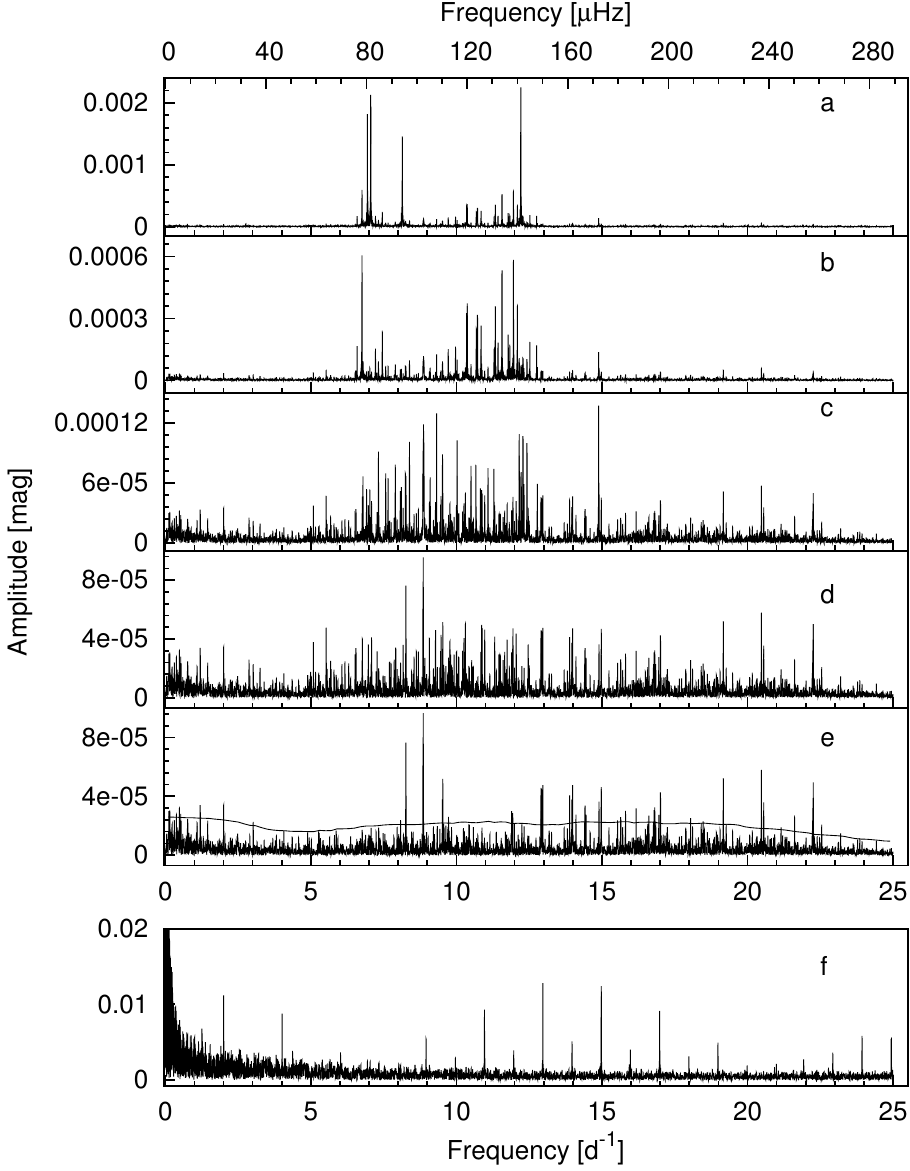}
 \caption{Amplitude spectrum at different steps of the frequency search process: 
{\bf (a)} original spectrum, {\bf (b), (c), (d), (e)} show the spectrum 
after subtracting 4,25,55 and 115 peaks, respectively. In panel {\bf (e)} the significance level is also given by continuous line. Panel {\bf (f)} shows the fine structure of the spectral window, where its y-axis is normalised to unity instead of mag.
 }
\label{fig-spec}
\end{figure}

We used the generally applied software packages for frequency analysis: SigSpec \citep{Reegen07}, Iterative Sine-wave Fitting (ISF) \citep{Vanicek71}, MuFrAn \citep{Kollath90}, and Period04 \citep{Lenz05}. Each of them has an advantage in some respect. SigSpec is fully automated and well-suited for the analysis of large light-curve samples and a fast, first step determination of the frequency content. ISF, MuFrAn, and Period04 provide iterative multi-frequency solutions by human interaction for a limited number of frequencies. 

The user-friendly graphical display of MuFrAn allows investigating the possible amplitude increase due to unresolved, closely spaced peaks. Period04 facilitates the error estimation and the determination of the S/N, which can be readily used as a stop criterion for the iterative frequency search.

Although SigSpec retained about 740 peaks until the significance limit of six \citep{Reegen07} was reached, and the consecutive highest amplitude solution of Period04 revealed 250 peaks until the S/N=4.0 limit \citep{Breger93}, we accepted only 115 frequencies in our final list. Our argument is based on the fine structure of the spectral window presented in panel {\bf (f)} of Fig\,\ref{fig-spec}. Note that the y-axis of the spectral window is normalised to unity, while the axis label refers to panels {\bf (a)} to {\bf (e)}. Of course, due to the high duty cycle, there are no high amplitude aliases. The obvious alias structure, connected to the orbital motion of the spacecraft, is lower than 2\%. In the range from 0 to 6\,d$^{-1}$ there is an alias structure with  decreasing amplitude. At any step of finding the next peak (prewhitening), the alias structure influenced (due to uncertainties) our next choice.

We omitted peaks supposedly connected to the fine structure of the spectral window. We may have missed some of the low amplitude frequencies of the pulsation, but not including spurious peaks in our final frequency list outweighed the loss of low amplitude frequencies. We omitted closely spaced peaks with frequency difference obviously smaller than the Rayleigh resolution of 0.0085\,d$^{-1}$. We further omitted frequencies with bigger difference than the Rayleigh resolution, where including both components in the solution artificially increased the amplitude of the first frequency in the least-square fit process. The different software packages resulted in the same solution concerning the frequencies with amplitudes higher than 0.05\,mmag. 

The different steps of the period searching process are shown in 
Fig.\,\ref{fig-spec}. For better visibility, only the 0 - 25~d$^{-1}$ range is 
displayed in the panels, since there are only spurious peaks (connected to the orbital period of CoRoT)
above the noise level out of this range. We also present the low-frequency region, since any tidally excited g-mode should be expected here.  Panel {\bf (a)} gives the original spectrum dominated by peaks with almost the same amplitude (12.21284, 7.05419 and 6.94259 d$^{-1}$) and a fourth one with slightly lower amplitude (8.14077\,d$^{-1}$).
Panels {\bf (b)}, {\bf (c)}, {\bf (d)}, and {\bf (e)} show the spectrum after
subtracting 4, 25, 55, and 115 frequencies, respectively. In panel {\bf (c)} a peak at 14.8846 d$^{-1}$ has the largest amplitude. Although it falls in the fine structure of the +1\,d$^{-1}$ alias of the CoRoT orbit frequency, the relatively high amplitude of this peak suggests that it should be included in our frequency list as an intrinsic real frequency. The other peaks in panel {\bf (c)} and {\bf (d)} show the large number of lower amplitude frequencies.

The low frequency part of both panels shows an increasing noise level, indicating that the subtracted binary light curve was not completely appropriate. In a later stage of the analysis, we compared it to the theoretical orbital light curve (Sec.\,\ref{sec-orbit}) that was much smoother than we had subtracted above (Sec.\,\ref{ssec-freqanal-data}). There was no change in the frequencies of neither the p-mode nor the g-mode region. Comparing the low frequency regions of panels {\bf (d)} and {\bf (e)}, two low amplitude frequencies deserve attention. One of the frequencies at  3.253(4) d$^{-1}$ ,  just above the S/N=4.0 significance level, may be interpreted as a possible tidally excited g mode. It is near an integer multiple of the  orbital period ($f_{111}$ = 108.98*$f_{orb}$). The other low frequency is 2.873(7)\,d$^{-1}$, which is not an integer multiple of the orbit frequency.

Panel {\bf (e)} presents our residual spectrum. The significance level is given by the continuous line. Period04 was used to obtain the significance value in each bin of 5\,d$^{-1}$, which was shifted by 2\,d$^{-1}$.  
The significance level (i.e., four times the mean amplitude of the noise) has a minimum of 15.8 ppm
at about 5 d$^{-1}$. 
Some of the residual peaks are above the significance level. These peaks (8.2631, 8.8597, 8.8468 and 9.5295\,d$^{-1}$) are closely spaced to one of our frequencies in our final list. For this reason we cannot resolve them without amplitude distortion. The 8.84687\,d$^{-1}$ peak is especially interesting, since two other close peaks (at 8.8682(8) and 8.8818(8)\,d$^{-1}$) are included in our final 
frequencies. The three peaks are not equidistant, but their spacing corresponds to a splitting with 46.7 - 73.5\,d. 
The peaks around 1,2,13,14,15, and 17\,d$^{-1}$ could be the remnants of the fine structure of CoRoT's orbital period aliases. Four peaks at 19.169, 20.4758, 20.5591, and 22.2555\,d$^{-1}$ are questionable. They are in the region of the $f_i+f_j$ linear combinations. It is possible to numerically find a close solution, such as $f_1+f_3$ = 19.155\,d$^{-1}$ or $f_1+f_2$ = 19.267\,d$^{-1}$, but they are out of the Rayleigh limit to the 19.169\,d$^{-1}$ that we found. The frequency  $f_{114}$ = 5.158\,d$^{-1}$ could be interpreted as $f_1-f_2$. Linear combinations would show that the frequencies involved belong to the same component. We can hardly expect linear combination of the other frequencies due to their low amplitudes. We decided not to include the questionable peaks in our final list. 

A special test was made to find a useful criterion when to stop the iterative process of the frequency analysis. Close frequencies (33 at around 12.212\,d$^{-1}$, 29 around 7.054\,d$^{-1}$, 14 around 8.140\,d$^{-1}$, and 25 around 10.369\,d$^{-1}$) were used to generate synthetic data on the whole timebase without adding any noise. The frequency analysis of the synthetic data by Period04 returned the input frequencies. However, in each case, groups of low amplitude but significant ``extra'' frequencies were found, although the input frequencies and amplitudes were constant in these cases (variability of frequencies and amplitudes can cause a similar effect). The test showed it was more secure to stop above the traditional significance limit. A mathematical approach of such problems in the treatment of space data was recently published by \citet{Balona14}.

Table\,\ref{tabl-freq} lists the first 55 frequencies, their amplitude (A$>$0.05 mmag), phases and signal-to-noise ratio value. The formal errors were calculated by Period04. The first uncertain digits are given in parentheses. The other frequencies up to 115 in the same format are given only electronically (Table\,\ref{tab-online}).  

The fits in Fig.\,\ref{fig-light} with four (black dashed line)
and 115 (white continuous line) frequencies nicely show that a part of the light curve (middle panel)
 could be remarkably well fitted by the four dominant modes, but another part 
of the light curve (bottom panel) shows large deviations from the measured values. The 115 frequencies perfectly fit the complicated light curve of \object{HD\,51844}\ shown by the reduction of the variances of the residuals (85\% with the four frequency fit and 98.8\% with the 115 frequency solution). The excited modes are distributed in the 5-15 d$^{-1}$ range, giving a constraint for seismic modelling.

\subsubsection*{Regularities in the frequency content}\label{ssec-resonance} 

The light-time effect due to the orbital motion appears as frequency modulation.   
In the case of an elliptical orbit, a multiplet structure appears \citep{Shibahashi12}. The side peak structure around the four highest amplitude modes has been discussed in \citet{Hareter14}, giving the values of the first and second order side-peaks, their amplitudes, and phases on the SigSpec solution.

Our final list of frequencies concentrated only on the doubtless, higher amplitude modes and contains only the triplet structure of the frequency modulation. No further triplets were found.
However, we found doublets with frequency spacings corresponding to an integer number (2,3,4,6,7,10) times of the orbital frequency (for example $f_{2}$ -- $f_{5}$ = 10 $f_{orb}$).
Thirty frequencies are in doublets (26\%). Including multiplets, 36.5\% of the frequencies are separated by integer multiples of the orbital frequency.

The large frequency separation is defined as the frequency spacing of consecutive radial orders for pulsators in the asymptotic regime, and is proportional to the inverse of the sound travel time. 
For low-order p mode pulsators ($\delta$ Sct\ stars), this separation becomes fuzzy in particular for evolved stars.  For non-radial ($\ell > 0$) modes, this separation is destroyed due to avoided crossing and mixed modes. However, the radial modes ($\ell = 0$)  approximately keep the separation (although not exactly); thus, it is justified to look for such a frequency spacing.

Two p-modes are multiples of the orbital frequency as in KIC 45444587
\citep{Hambleton13}. The ratios of the p modes to the orbital frequencies are
$f_{22}$/$f_{orb}$ = 333.999849 and $f_{36}$/$f_{orb}$ = 318.9922410. We conclude that these p-modes can be tidally excited modes. Although it would be straightforward to suppose that the well-separated groups around $f_1$ belongs to one component and around $f_2$ to the other component, we did
not find any argument for it. The regularities connect the two groups, like
the possible large separation (spacing between the consecutive radial orders), which are
around 2.0 -- 2.2\,d$^{-1}$ ($f_{8} - f_{4}$ = 2.23, $f_1 - f_{22}$ = 2.24, $f_1 - f_4$ = 2*2.04). A slightly
different value is shown between $f_1 - f_2$ (2* 2.58), which does not fit the previous
sequence. The frequency  $f_6$ could be the rotational split of $f_1$ with 0.2538 d$^{-1}$, which agrees
with the rotational period of secondary component (Sec.\,\ref{sec-orbit}). There is no
rotational split around $f_2$ with the previous value but $f_5$ differs by about
10*$f_{orb}$ (0.299\,d$^{-1}$), which is in the range of the rotational velocities of the components (Sec.\,\ref{sec-orbit}). The possible large separation 2.0 -- 2.2\,d$^{-1}$ could constrain the seismic modelling.

A 3:2 resonance was immediately noticed between $f_1$ = 12.21284(4) and $f_4$ = 8.14077(3) d$^{-1}$ frequencies. Normally, the p-modes excited in a single $\delta$ Sct\ star do not show such a frequency ratio except if subharmonics are also excited due to non-linear effects.
To see the whole situation, a systematic search was done on the ratio of all frequencies to all frequencies. Resonances were accepted if the frequency difference between the real frequency and the calculated frequency with the exact resonance value were lower than the Rayleigh resolution. Altogether, 32 resonance pairs were noticed with 53 independent frequencies that were included out of the 115 frequencies listed in the electronic table. Different kinds of frequency relations were found including one first (2:1) harmonic, one second (3:1) harmonic, and a 2:5 resonance. We also searched for 3:2, 3:4 and 3:5 resonances. The resulting list of resonances is given in Table\,\ref{tab-group}. Altogether we found 9, 11, and 8 pairs with 3:2, 3:5 and 3:4 resonances, respectively. Our conclusion is that about half of the peaks in the $\delta$ Sct\ p-mode region do not correspond to independently excited modes but are excited by resonances.

\begin{table}
 \caption{Resonance pairs. The first two columns give the labels of the frequencies included in our final list. The third column gives the frequency difference between the real frequencies (second column) and the calculated ones (first column) with the exact resonance values.}
 \label{tab-group}
 \centering
 \begin{tabular}{rrr}
  \hline\hline
Freq. & Freq. & Difference  \\
(d$^{-1}$) & (d$^{-1}$)  & (d$^{-1}$)  \\
  \hline
\multicolumn{3}{c}{3:2} \\
  \hline
$f_{87}$ & $f_{26}$ & 0.00781 \\ 
$f_{4}$  & $f_{1}$  & 0.001685 \\
$f_{54}$ & $f_{30}$ & 0.008690 \\
$f_{35}$ & $f_{62}$ & 0.009810 \\
$f_{52}$ & $f_{8}$  & 0.00099 \\ 
$f_{90}$ & $f_{71}$ & 0.00195 \\ 
$f_{101}$ & $f_{48}$ & 0.00444 \\
$f_{113}$ & $f_{80}$ & 0.000050\\
$f_{109}$ & $f_{88}$ & 0.00215 \\ 
  \hline
\multicolumn{3}{c}{3:5} \\
  \hline
$f_{45}$ & $f_{20}$ & 0.00112 \\ 
$f_{15}$ & $f_{34}$ & 0.001547 \\ 
$f_{66}$ & $f_{108}$ & 0.006533 \\ 
$f_{3}$ & $f_{7}$ & 0.004854 \\    
$f_{46}$ & $f_{67}$ & 0.000083 \\ 
$f_{19}$ & $f_{62}$ & 0.00715 \\ 
$f_{90}$ & $f_{57}$ & 0.00300 \\ 
$f_{102}$ & $f_{38}$ & 0.0000 \\ 
$f_{97}$ & $f_{103}$ & 0.00733 \\
$f_{96}$ & $f_{65}$ & 0.00733 \\
$f_{96}$ & $f_{94}$ & 0.00206 \\
  \hline
\multicolumn{3}{c}{3:4} \\
  \hline
$f_{27}$ & $f_{34}$ & 0.00423 \\ 
$f_{64}$ & $f_{10}$ & 0.00658 \\ 
$f_{85}$ & $f_{86}$ & 0.006533 \\
$f_{4}$ & $f_{13}$ & 0.000036 \\ 
$f_{54}$ & $f_{51}$ & 0.00563 \\ 
$f_{100}$ & $f_{70}$ & 0.00690 \\
$f_{45}$ & $f_{79}$ & 0.007867 \\
$f_{96}$ & $f_{43}$ & 0.001087 \\
  \hline
 \end{tabular}
\end{table}

\section{Spectroscopy}
\label{sec-spec}

\begin{table*}
\caption{Observation log of the spectroscopy.}             
\label{tab-obslog}      
\centering                          
\begin{tabular}{c c c c c c c}        
\hline\hline                 
HJD & label & Integration time & S/N & orbital phase\tablefootmark{1} & spectrograph & Observer \\    
\hline                        
   2455178.76490 & 602 & 1400 s & 199.0 &  0.477  & HARPS & M. Rainer\\      
   2455192.84634 & 801 & 1200 s & 174.4 &  0.897  & HARPS & M. Hareter\\
   2455194.86688 & 831 & 1200 s & 137.9 &  0.958  & HARPS & M. Hareter\\
   2455195.85916 & 820 & 1200 s & 154.1 &  0.987  & HARPS & M. Hareter\\
\hline   
   2456375.35553 &  23 & 1200 s &  69.8 &  0.198  & HERMES & P. Marcos-Arenal\\  
   2456376.37884 &  24 & 1200 s &  97.3 &  0.229  & HERMES & P. Marcos-Arenal\\  
   2456377.3494  &  25 & 1200 s & 101.2 &  0.258  & HERMES & P. Marcos-Arenal\\  
   2456380.35516 &  28 & 1200 s &  97.6 &  0.348  & HERMES & P. Marcos-Arenal\\  
   2456381.35728 &  29 & 1200 s &  91.4 &  0.377  & HERMES & P. Marcos-Arenal\\  
   2456382.35234 &  30 & 1200 s & 108.2 &  0.407 & HERMES & P. Marcos-Arenal\\  
   2456383.35062 &  31 & 1200 s &  93.2 &  0.437  & HERMES & P. Marcos-Arenal\\  
   2456384.35832 &  32 & 1600 s &  91.6 &  0.467 & HERMES & J. Vos\\  
   2456388.3539  &  36 & 1400 s & 102.2 &  0.586  & HERMES & J. Vos\\  
   2456389.35639 &  37 & 1400 s &  79.5 &  0.616  & HERMES & J. Vos\\  
   2456390.36529 &  38 & 1200 s &  61.0 &  0.646  & HERMES & J. Vos\\  
   2456391.35329 &  39 & 1200 s &  98.9 &  0.676  & HERMES & J. Vos\\  
   2456392.3528  &  40 & 1000 s & 108.8 &  0.706  & HERMES & J. Vos\\  
   2456393.36444 &  41 & 1200 s & 108.2 &  0.736  & HERMES & J. Vos\\  
   2456394.37935 &  42 & 1600 s &  89.7 &  0.766  & HERMES & P. Lampens\\  
   2456395.37624 &  43 & 2000 s &  84.2 &  0.796 & HERMES & P. Lampens\\  
   2456396.38563 &  44 & 1500 s & 103.7 &  0.826 & HERMES & P. Lampens\\  
   2456397.3794  &  45 & 1800 s &  94.6 &  0.856  & HERMES & P. Lampens\\  
   2456398.37166 &  46 & 1800 s &  94.4 &  0.885  & HERMES & P. Lampens\\  
\hline
   2456378.60500 &  Mc26 & 2x1200 s &  156 &  0.292  & McDonald & P. De Cat\\  
   2456379.60590 &  Mc27 & 2x1200 s &  130 &  0.322  & McDonald & P. De Cat\\  
   2456380.60990 &  Mc28 & 2x1200 s &  131 &  0.352  & McDonald & P. De Cat\\  
   2456381.60630 &  Mc29 & 2x1200 s &  81 &  0.382  & McDonald & P. De Cat\\
   2456383.60520 &  Mc31 & 2x1200 s &  - &  0.441  & McDonald & P. De Cat\\
\hline    
\end{tabular}
\tablefoot{
\tablefoottext{1}{Phase zero-point: JD 2\,456\,368.712 }  
}                       
\end{table*}
The periodic brightening event prompted us to organise a spectroscopic campaign with the aim to monitor the radial velocities of the components during one orbital cycle. The coverage was, however, not sufficient to do mode identification of the pulsation modes. The HARPS spectra from late 2009, as obtained during LP185.D-0056, already indicated a composite spectrum, but, at that time, it was not clear how many components are present.

Table\,\ref{tab-obslog} gives the observing log. The HJD refers to the time of mid-exposure, the signal-to-noise ratio was estimated between 5805 and 5825\,\AA, except for the McDonald spectra, where it was estimated near 4690\,\AA. Note that the HARPS spectra are not numbered in chronological order.

The HARPS spectra were reduced using a semi-automatic pipeline
developed at INAF-OAB, which gives two different outputs:  1) normalised and
merged spectra, and 2) both normalised and un-normalised spectra with the
orders separated. In both cases, the spectra were flat-field corrected,
de-blazed using the continuum of a hot star, wavelength calibrated, and
the barycentric correction was applied. 

The HERMES spectra were reduced in an automatic way with the dedicated HERMES pipeline, which provides un-normalised merged spectra as a function of wavelength.
The McDonald Spectra were reduced using IRAF\footnote{IRAF is distributed by the National Optical Astronomy Observatories, which are operated by the Association of Universities for Research in Astronomy, Inc., under cooperative agreement with the National Science Foundation.}.
The normalisation was done by fitting low-order polynomials to continuum windows in the data. We took care  not to distort the hydrogen lines.

\begin{figure}
   \centering
   \includegraphics[width=\hsize]{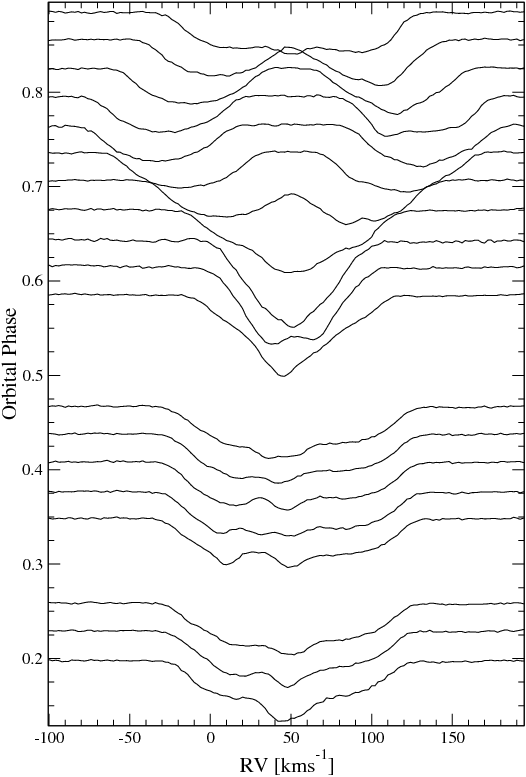}
      \caption{LSD profiles of the HERMES spectra phased with the orbital period. The phase 0 corresponds to JD 2\,456\,368.712.}
         \label{fig-hermes-lsd}
   \end{figure}

\begin{figure}
   \centering
   \includegraphics[width=\hsize]{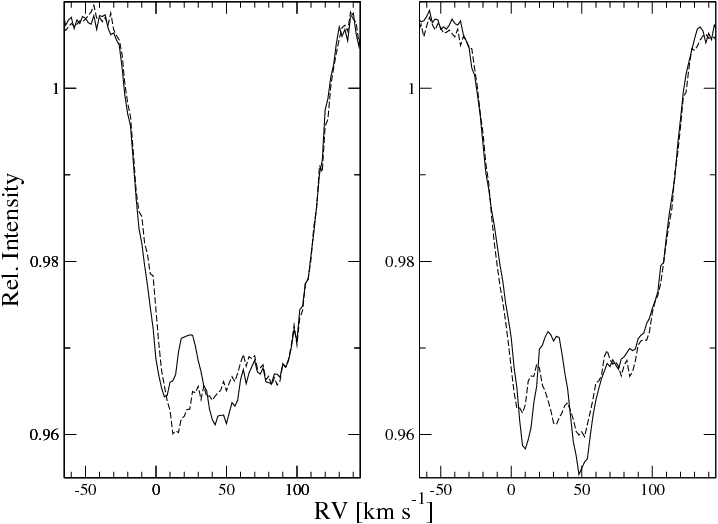}
      \caption{LSD profiles of two McDonald spectra taken on March 28th, 2013 (phase 0.352) consecutively (left panel, full line: first exposure, dashed line second exposure) compared to two LSD profiles of two spectra from HERMES (March 28, 2013, phase 0.348, full line and March 29, 2013 phase 0.377, dashed line).}
         \label{fig-mcd-lsd}
   \end{figure}

Figure\,\ref{fig-hermes-lsd} shows the least squares deconvolution (LSD) profiles phased with the orbital period. One component shows a smooth rotation profile, while the other shows strong line profile variation. This situation is best visible in the profiles around the phase of 0.75 and at the phases near quadrature (i.e., phases 0.4 and 0.8). These profiles show that only the component, which has redshifted lines at phases near 0.75 shows either non-radial pulsation or a close companion. To examine this issue, LSD profiles from consecutive McDonald observations were compared (see Fig.\,\ref{fig-mcd-lsd}). The McDonald spectra are separated by 20 minutes, thus, the fast change implies that the line profile variation is caused by pulsation rather than by the close binary scenario.

\subsection{Element abundances}
\label{sec-abund}

\begin{figure}
   \centering
   \includegraphics[width=\hsize]{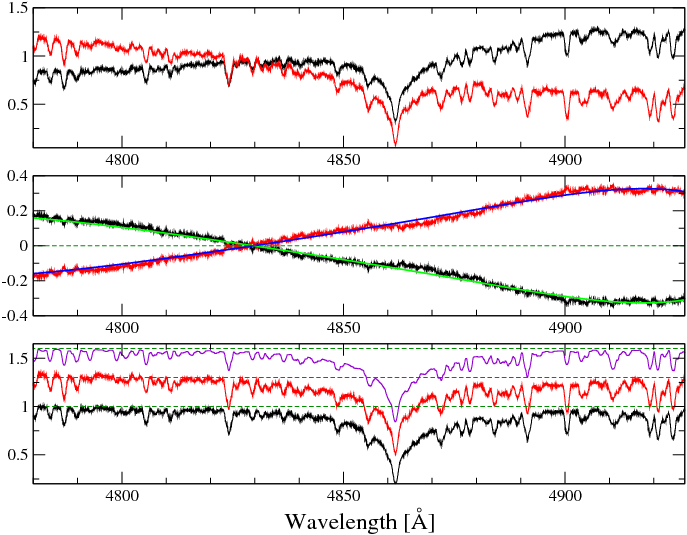}
      \caption{Disentangled spectra (top panel), bias (middle panel) and polynomial fits to the bias (smooth lines), and corrected spectra (bottom panel). The spectra shown in black and red correspond to component A and B, respectively. For the corrected spectra the fits to the bias were simply subtracted. The top spectrum of the bottom panel is a synthesis calculated with T$_{\mathrm{eff}}$ = 6800\,K, $\log g$ = 3.2 and solar metallicity (violet line). The dashed horizontal lines indicate the continuum. The individual spectra are shifted for better visibility. A colour version of this figure is available online.}
         \label{fig-disent}
\end{figure}

\begin{figure}
   \centering
   \includegraphics[width=\hsize]{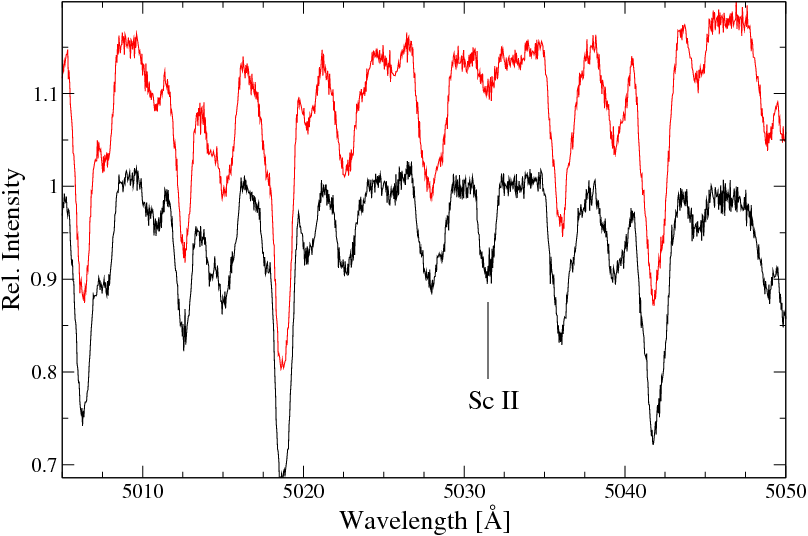}
      \caption{Disentangled spectra showing the clearly different strengths of \ion{Sc}{II} line at 5031\AA. The spectrum of the pulsating component is displayed on top the spectrum of the non-pulsating component below. No attempt to re-normalise the spectra was done.}
         \label{fig-disent-sc}
\end{figure}

We present the results of the abundance analysis based on spectrum \#43 of the HERMES series and spectrum \#602 and \#831 of the HARPS series. These spectra were taken at different orbital phases, namely at 0.796, 0.477 and 0.958, respectively.
The goal of the abundance analysis presented here is to confirm or reject the $\delta$ Del classification and to derive the fundamental parameters. The HERMES spectra \#40 - 44 cover the phase of maximum separation of the spectral lines of the two components. The spectrum \#43 was chosen, because the lines of the components are widely separated; the HARPS spectra \#602 and \#831 were selected, because they are blended and they provide a superior S/N over the HERMES spectra. The consistency of the abundances derived from spectrum \#43 was assessed using the mentioned HARPS spectra.

We followed the spectrum synthesis approach to determine the abundances. The ATLAS9 model atmospheres \citep{Kurucz93a} and line lists from VALD \citep{Piskunov95, Kupka99, Ryab97} were used. The synthetic spectra were calculated using the synth3 code \citep{Kochukhov07}. To fit the spectral lines, the visualisation software package BinMag3\footnote{url:http://www.astro.uu.se/$\sim$oleg/} was used, which conveniently allows fitting abundances of synthetic spectra to observations. To be able to use this feature, the luminosity ratio plays an important role. The hydrogen lines suggest almost identical components, the luminosity ratio was initially set to 1.0.   

To determine the Fe abundances, we used only lines of spectrum \#43, where the lines were not blended. In this way, we could use 25 \ion{Fe}{I} lines and 7 \ion{Fe}{II} lines per star. The microturbulent velocity (v$_\mathrm{mic}$) was determined by minimising the standard deviations of the \ion{Fe} {I} abundances for different v$_\mathrm{mic}$ ranging from 2.0 to 5.0\,km\,s$^{-1}$. For the component showing no LPV (hereafter component A) v$_\mathrm{mic}$ = 3.8\,km\,s$^{-1}$ yielded the minimum standard deviation while the star with the strong LPV (hereafter component B) showed the minimum standard deviation of the \ion{Fe} {I} abundances at v$_\mathrm{mic}$ = 2.6\,km\,s$^{-1}$. These results were checked with the HARPS spectra \#602 and \#831.

The effective temperatures were determined by fitting synthetic spectra to the observed H$\beta$ line and were subsequently refined by using the excitation potential vs. abundance. For component B, the initial guess T$_{\mathrm{eff}}$ = 6800\,K  was correct, while the temperature of component A had to be corrected by 400\,K (T$_{\mathrm{eff}}$ = 7200\,K).

The ionisation equilibrium could only be achieved if low $\log g$ values were allowed. The imbalance of the abundances from \ion{Fe}{I} and \ion{Fe}{II} increased for higher surface gravities. Even for $\log g$ = 3.5, the ionisation equilibrium was not achieved. Therefore we adopted $\log g$ = 3.2 for both components. The uncertainties of the temperatures were estimated to 200\,K and for $\log g$ to 0.2 dex.

The ratio of the radii was estimated by visually inspecting the cores of the hydrogen lines comparing to synthetic binary spectra. Radius ratios of 1 fit best, while the syntheses did not fit the observations any more below 0.9 and above 1.1. Thus, we adopted 1.0 $\pm$ 0.1 for the radius ratio. To fit synthetic line profiles to the observed line profiles, we scaled the observations accordingly. 

The abundances of additional 20 elements could be derived by this method (see Table\,\ref{tab-abundances}). The chief difference of the two stars is the Sc abundance, which is deficient in the non-radially pulsating component by $\approx$ 1 dex and slightly underabundant in the other component. The Y and Ba are overabundant by 1 dex, where the non-radially pulsating component shows a larger Y abundance than component A. Despite the large uncertainties of the abundances of the s-process elements, the qualitative results are valid and show a clear Am-pattern for both components. 

A successful disentangling of spectra requires a homogeneous phase coverage. The attempt to disentangle them using the code {\sc fdbinary} \citep{Ilijic03} led to undulations and distortions of the H$\beta$ wing (around 4875\AA, Fig.\,\ref{fig-disent}). The bottom panel shows the re-normalised spectra, where the bias (solid lines middle panel) was subtracted. Therefore, we did not use the disentangled spectra for the determination of the abundances. 
Nevertheless, the disentangling confirmed that both components
are very similar and are compatible with T$_{\mathrm{eff}}$ and $\log g$ that are determined on the composite spectra. A synthetic spectrum with solar metallicity is shown for comparison in the bottom panel of Fig.\,\ref{fig-disent}. Note the strong Y lines at 4900\,\AA.
The disentangled spectra also confirm different line strengths of the Sc lines of both components, which is most obvious at the \ion{Sc}{II} line 5031\AA\ (shown in Fig.\,\ref{fig-disent-sc}). 

\begin{table}
\caption{Elemental abundances ($\log N_X/N_{tot}$) compared to the sun's abundances; the first number in brackets is the standard deviation and the numbers of lines are separated by '';''.  The solar abundances are taken from \citet{Asplund09}. The abundances of the elements below the line are derived by fitting line blends. }             
\label{tab-abundances}      
\centering                          
\begin{tabular}{c c c c }        
\hline\hline                 
Element & Component A & Component B & Sun \\
\hline
Fe   &  -4.85 (0.3;28) & -4.44 (0.2;28) &  -4.54 \\
Ca   &  -5.67 (0.18;8) & -5.70 (0.27;6) &  -5.70 \\
Ni   &  -5.82 (0.19;5) & -5.44 (0.16;6) &  -5.82 \\
Mg   &  -4.43 (0.08;3) & -4.70 (0.30;4) &  -4.44 \\
Ti   &  -7.60 (0.37;3) & -7.40 (0.55;3) &  -7.09 \\
Ba   &  -8.76 (0.33;3) & -8.90 (0.13;3) &  -9.86 \\
Y    &  -9.33 (0.36;3) & -8.98 (0.54;2) &  -9.83 \\
Sc   &  -9.10 (-;1)    & -10.05 (0.25;3) & -8.89 \\
Si   &  -4.60 (0.16;3) & -4.35  (0.23;2) & -4.53 \\
S    &  -4.73 (0.01;2) & -4.85 (0.07;2) & -4.92 \\
Mn   &  -7.00 (0.14;2) & -6.61 (-;1)  & -6.61 \\
O    & -3.27 (-;1)     & -3.35 (-;1) & -3.35\\
C    & -4.11 (-;1)     & -3.85 (-;1) & -3.61 \\
Li   & -9.50 (-;1)     & -9.50 (-;1) & -10.99 \\
\hline
La   & -9.94 (-;1)  & -9.94 (-;1) & -10.94 \\
Ce   & -9.46 (-;1) & -9.46 (-;1) & -10.46 \\
Sr   & -8.67 (-;1) & -8.17 (-;1) & -9.17\\  
Zr   & -8.46 (-;1) & -8.96 (-;1) & -9.46 \\
Nd   & -10.62 (-;1) & -9.62 (-;1) & -10.62 \\
Eu   & -10.52 (-;1) & -10.52 (-;1)& -11.52 \\ 
Lu   & -11.0 (-;1) & -10.44 (-;1) & -11.94 \\

\hline
\end{tabular}
\end{table}

\subsection{Radial velocities}
\label{ssec-rv}

\begin{figure}
   \centering
   \includegraphics[width=\hsize]{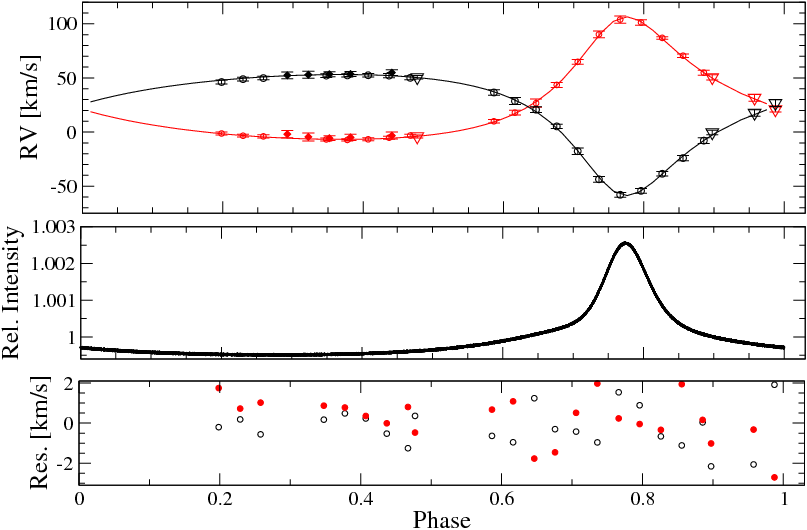}
      \caption{Radial velocity curves (upper panel) showing the measurements (symbols with error bars) and the fitted RV solutions (full lines). The open circles, filled squares and the open triangles represent the HERMES, McDonald and HARPS spectra, respectively. The phased binary light curve is shown in the middle panel for comparison. The bottom panel shows the residuals in km\,s$^{-1}$ for component A (open circles) and B (filled circles).}
         \label{fig-hd5-rvcurve}
   \end{figure}

The radial velocities and the projected rotational velocities were determined with the software BinMag3. This code allows fitting synthetic composite spectra to the observed composite spectra.
The $v\sin i$ of the two components was determined by using the HERMES spectra \#40 to 45, where the lines of the two components were clearly resolved. We used 7 to 12 individual measurements in different wavelength regions for each spectrum and averaged all measurements. The resulting $v\sin i$ is equal for both components: 41.37$\pm$1.3 and 41.73$\pm$1.7\,km\,s$^{-1}$, respectively. 

To determine the radial velocities, we used the same tool by keeping the $v\sin i$ fixed at the mentioned average values and allowing the adjustment in RV only. About 16 regions in each individual spectrum were selected for fitting the RVs. In total, about 420 individual measurements were performed. The RV measurements with their corresponding standard deviations are given in Table\,\ref{tab-rvs}. The systemic velocity ($\gamma$ velocity) was not subtracted; it was determined in the following analysis. The resulting RV curve is shown in Fig.\,\ref{fig-hd5-rvcurve} (upper panel), with the fits to the RVs of both components. The binary light curve (see Sec.\,\ref{sec-orbit}) derived from the CoRoT photometry is shown for comparison in the middle panel.
The phase of maximum RV difference for the primary component coincides with the phase of maximum brightening, which indicates that the brightening is in fact due to the irradiation/reflection effect at periastron. Furthermore, the radial velocities of both components are of the same magnitude, which means nearly equal masses. This is another evidence of both components being nearly identical stars.

We used the T$_{\mathrm{eff}}$, $\log g$, $v\sin i$, RVs, and the ratio of the radii as input for the orbital modelling. The abundances were used as input for the seismic modelling (Sec.\,\ref{sec-model}).

\begin{table}
\caption{Radial velocities derived from individual spectra. The errors are the standard deviations of about 16 individual measurements.}             
\label{tab-rvs}      
\centering                          
\begin{tabular}{c c c c }        
\hline\hline                 
 Label &  BJD-2451545.0 &  RV$_A$[km\,s$^{-1}$] &  RV$_B$ [km\,s$^{-1}$] \\
\hline 
 602 &  3633.76490 &   50.61$\pm$1.53 & -4.11$\pm$1.35 \\ 
 801 &  3647.84634 &  -0.38$\pm$2.05 & 50.59$\pm$2.46 \\ 
 831 &  3649.86688 &  17.65$\pm$2.06 & 31.54$\pm$3.07 \\ 
 820 &  3650.85916 &  26.65$\pm$3.13 & 20.86$\pm$2.29 \\ 
\hline
 23 &  4830.35553 &  46.21$\pm$1.76  & -1.33$\pm$1.30 \\ 
 24 &  4831.37884 &  48.89$\pm$1.75  & -3.40$\pm$1.29 \\ 
 25 &  4832.34940 &  49.87$\pm$1.63  & -3.94$\pm$1.46 \\ 
 28 &  4835.35516 &  52.17$\pm$1.64  & -6.85$\pm$1.48 \\ 
 29 &  4836.35728 &  52.34$\pm$1.49 & -7.25$\pm$1.01  \\ 
 30 &  4837.35234 &  52.48$\pm$1.71 & -6.71$\pm$1.41 \\ 
 31 &  4838.35062 &  52.14$\pm$1.89 & -5.24$\pm$1.45 \\ 
 32 &  4839.35832 &  50.09$\pm$1.83 & -3.26$\pm$1.25\\ 
 36 &  4843.35390 &  36.56$\pm$2.51 & 9.95$\pm$1.64\\ 
 37 &  4844.35639 &  28.62$\pm$3.20 &  17.90$\pm$2.16 \\ 
 38 &  4845.36529 &  20.65$\pm$2.65 & 26.67$\pm$3.78 \\ 
 39 &  4846.35329 &   5.21$\pm$1.98 & 43.52$\pm$2.22 \\ 
 40 &  4847.35280 &  -17.71$\pm$3.09 & 64.86$\pm$2.19 \\ 
 41 &  4848.36444 &  -43.66$\pm$2.63 & 90.20$\pm$3.12 \\ 
 42 &  4849.37935 &  -57.95$\pm$2.18 & 103.93$\pm$3.38 \\ 
 43 &  4850.37624 &  -54.35$\pm$2.13 & 101.22$\pm$2.57 \\ 
 44 &  4851.38563 &  -38.53$\pm$2.00 & 87.04$\pm$1.16 \\ 
 45 &  4852.37940 &  -24.03$\pm$2.42 & 70.50$\pm$1.51 \\ 
 46 &  4853.37166 &  -7.92$\pm$2.62  & 54.84$\pm$2.28 \\ 
\hline
 Mc26 &  4833.60500 &  52.44$\pm$3.10 &  -1.97$\pm$3.36 \\ 
 Mc27 &  4834.60590 &  52.96$\pm$3.29 &  -4.54$\pm$3.66 \\ 
 Mc28 &  4835.60990 &  53.41$\pm$2.41 &  -5.67$\pm$2.17 \\ 
 Mc29 &  4836.60630 &  53.61$\pm$2.35 &  -4.93$\pm$2.87 \\ 
 Mc31 &  4838.60520 &  54.88$\pm$2.70 & -3.26$\pm$3.22 \\ 
\hline
\end{tabular}
\end{table}

\section{Orbital parameters}
\label{sec-orbit}

\begin{figure*}
\fontsize{12}{12}\selectfont
\includegraphics[width=84mm]{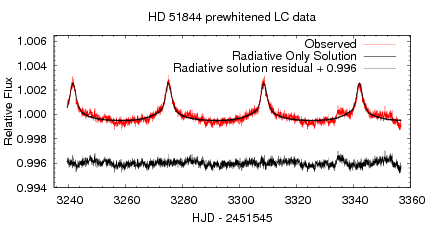}\includegraphics[width=84mm]{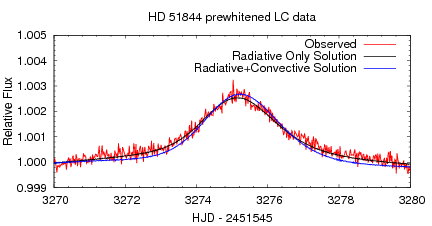}
 \caption{Binary light curve solutions of \object{HD\,51844}. {\it Left panel:} The binary light curve of \object{HD\,51844}\ after the removal of the pulsational variation (red) and the best solution when both stars have radiative atmosphere (black). The residual curve is also plotted in the bottom. {\it Right panel:} Comparison of the radiative solution to the 'best-fit' light curve solution for a convective secondary atmosphere.}
 \label{fig-lc_sol}
\end{figure*}

To derive the orbital parameters we used the cleaned light curve of Sec.\,\ref{sec-lightcurve}, 
that is prewhitened by frequencies in the range of 5-25\,d$^{-1}$, the radial velocities of
Sec.\,\ref{sec-spec}, and the parameters derived from spectroscopy.
Since no eclipses were observed, two minor effects of ellipsoidal light variation and the reflection or irradiation effect are visible in the light curve. Doppler-boosting, which might have a comparable amplitude to the previous ones \citep{Zucker07}, does not play any role in \object{HD\,51844}, 
due to the similarity of the two stars in both mass and brightness.

A simultaneous radial velocity and light curve analysis was performed (case of a detached binary) but did not lead to a unique solution. A variety of different solutions was possible, since the light curve does not allow constraining all of the input parameters. The RV curves are fitted almost exactly but the inclination was not at all constrained; thus, it could be obtained only from the light curve. 

The {\sc lightcurvefactory} \citep[see, e.g.,][]{Borkovits13} program allowed us to find a decent solution for both data types simultaneously when some of the parameters were fixed a priori.
When restricting our analysis to a single band light curve of a detached binary, this code was almost equivalent with the recent versions of the Wilson-Devinney (WD) \citep[see, e.~g.,][for recent improvements]{Wilson10} and PHOEBE codes \citep{Prsa05} with the added benefit of its flexibility in constraining the model parameters and the inclusion of Doppler-boosting effect (not used here). 

Our light curve modelling was based on the spectroscopic results ($T_{\rm{eff}}$, $\log g$ and $V_{\rm{rot}}\sin i_{\rm{rot}}$), and the orbital parameters  ($e$, $\omega$, $P_{\rm{orb}}$, $\tau$),  mass ratio ($q$), and projected mass of the components ($m\sin^3i_{\rm{orb}}$), which were obtained from the RV analysis. The $\log{g}$ values and the inclination ($i$) were the adjustable parameters. The $T_{\rm{eff,A}}$ = 7300\,K  (in the error bar of the spectroscopic value) was also fixed, since the light curve is sensitive to only the ratios of the temperatures. The adjustment of the $\log g$ values, which are observables of spectroscopic solution, is a speciality of the presently used program, giving opportunity to control the ratio of the radii and the orbital semi-major axis (i.e., the dimensionless fractional stellar radii) for the components through a directly observable spectroscopic parameter. The solution was  verified in each phase by PHOEBE. 

The fixing of $T_{\rm{eff,A}}$ was particularly critical in the present situation, as the stars were located near the transition region between convective envelopes and radiative atmospheres. Technically, this is reflected in a quick variation in the values of both the bolometric albedos and the gravity darkening exponents \citep{Claret98}. Their practically almost discontinuous nature, however, makes it possible to distinguish between the convective or radiative atmosphere solutions for the members of \object{HD\,51844}. 

The amplitude ratio of the reflection/irradiation effect to the ellipsoidal variation (consequently, the ``heartbeat'') strongly depends upon the bolometric albedos and gravity darkening exponents. Therefore, the net light-curve variation significantly differs from each other both in amplitude and shape. 

After the first preliminary runs we concluded that we can find reliable solutions only if both the stars have radiative envelopes. Therefore, the final parameter refinement was calculated only for the radiative case. In this stage, the orbital elements were also allowed to vary between the probable error limits of the RV solution. In Table\,\ref{tab-rv-orbitparam}, we list the orbital and stellar parameters.

Figure\,\ref{fig-lc_sol} shows the fit of the light curve using the radiative envelope scenario (left panel) and a comparison between the radiative and the convective envelope scenario for the secondary component. Thus, a hotter T$_{\mathrm{eff}}$ for component B results in a better fit to the light curve than the cooler T$_{\mathrm{eff}}$ from spectroscopy.

The orbital parameters enable us to calculate the theoretical Doppler shift of the pulsation frequencies. For the CoRoT light curve, we checked the five highest amplitude modes and found the first four highest amplitude frequencies have higher frequencies at apastron and lower frequencies at periastron in each orbit. Table\,\ref{tab-mod} gives the period differences in seconds from the apastron passage to the consecutive periastron passage and from the periastron passage to the consecutive apastron passage and so on. The differences and range from 1.0 -- 8.6 seconds. The shift of the frequencies is of the order of half of the Rayleigh resolution (0.0833\,d$^{-1}$). The expected Doppler shift of the periods was calculated according to

\begin{displaymath}
 f_{p} = f \left(1-\frac{K (1+e) \cos(\omega)} {c}\right), 
\end{displaymath}
\begin{displaymath}
 f_{a} = f \left(1+ \frac{K (1-e) \cos(\omega)} {c}\right), 
\end{displaymath}

\noindent The corresponding values of Table\,\ref{tab-rv-orbitparam} were inserted and for $f$ the corresponding frequency values. The consistency of the signs and the agreement with the theoretical Doppler shift with the resonances provides strong evidence that only the {\it secondary} component is pulsating.

The binary modelling modified the result of spectroscopy remarkably for
two parameters, the T$_{\mathrm{eff}}$ and $\log g$ of the components. The T$_{\mathrm{eff}}$ of the secondary 
(the pulsating component) was suggested to be hotter with 500 K. The $\log g$ of both components were increased by 0.4 dex, which is far above the error bars of the spectroscopic determination.

\begin{table}
 \caption{Orbital and stellar parameters derived from the binary light curve analysis.}
 \label{tab-rv-orbitparam}
 \begin{tabular}{@{}lll}
  \hline
\multicolumn{3}{c}{orbital parameters} \\
\hline
  $P_\mathrm{orb}$ (days) & \multicolumn{2}{c}{$33.4983\pm0.0020$} \\
  $a$ (R$_\odot$) & \multicolumn{2}{c}{$69.37\pm0.46$} \\
  $e$ & \multicolumn{2}{c}{$0.484\pm0.020$}  \\
  $\omega$ ($\degr$)& \multicolumn{2}{c}{$166.68\pm2.0$}  \\ 
  $i$ ($\degr$) & \multicolumn{2}{c}{$69.8\pm5.0$}  \\
  $\tau$ (BJD) & \multicolumn{2}{c}{$2\,454\,786.4223\pm0.25$} \\
  $K_{1}$ (km\,s$^{-1}$) & \multicolumn{2}{c}{$55.92\pm0.50$} \\
  $K_{2}$ (km\,s$^{-1}$) & \multicolumn{2}{c}{$56.63\pm0.50$} \\
  $\gamma$ (km\,s$^{-1}$) & \multicolumn{2}{c}{$23.36\pm0.23$} \\
  $q$ & \multicolumn{2}{c}{$0.988\pm0.012$} \\
  \hline  
\multicolumn{3}{c}{stellar parameters} \\
\hline
   & Primary & Secondary \\
  \hline
 $m$ (M$_\odot$) & $2.01\pm0.20$ & $1.97\pm0.19$ \\
 $R$ (R$_\odot$) & $3.69\pm0.18$ & $3.52\pm0.17$ \\
 $T_\mathrm{eff}$ (K)& $7300\pm200$ & $7300\pm200$ \\
 $L$ (L$_\odot$) & $34.1\pm0.4$ & $31.1\pm0.4$ \\
 $\log g$ (dex) & $3.62\pm0.01$ & $3.65\pm0.01$ \\
 $P_\mathrm{rot}$ (days) & $4.07\pm0.28$ & $3.89\pm0.29$ \\
 \hline

 \end{tabular}
\end{table}

\begin{table*}
 \caption{ Frequency modulation due to the orbital motion. The period changes between the apastron and periastron phases are indicated in seconds and the expected light time effect (Doppler shift) from the binary orbit solution is given in the last column. }
 \label{tab-mod}
 \centering
 \begin{tabular}{lllllll}
  \hline\hline
Freq. & ap1-per2 & per2-ap2 & ap2-per3 & per3-ap3 & ap3-per4 & expected \\
  \hline
(d$^{-1}$) & s & s & s & s & s & s \\
  \hline
12.209242 & +1.6 & -3.2 & +1.0 & -2.8 & +2.9 & $\pm$2.6\\
7.054777  & +3.8 & -4.6 & +4.9 & -5.4 & +5.4 & $\pm$4.5\\
6.944380  & +8.6 & -7.0 & +5.6 & -6.1 & +4.2 & $\pm$4.6\\
8.141873  & +2.9 & -6.5 & +5.2 & -1.7 & +3.1 & $\pm$4.2\\
  \hline
 \end{tabular}
\end{table*}

\section{Evolution and pulsation models}
\label{sec-model}

\begin{table*}
\caption{Parameters of models for component B, including the range of excited modes; the last column gives the mean density.}
\label{tab-models}
\centering
\begin{tabular}{cccccccc}
\hline\hline
Model & Mass    & T$_{\mathrm{eff}}$ & $\log g$     & age & range of excited modes & $\Delta\nu$ & $\overline{\rho}$\\
 No.  & [$M_{\sun}$] & [K]    & [cm/s$^2$] &  Gyr       & [d$^{-1}$] & [d$^{-1}$] & [g\,cm$^{-3}$] \\
\hline
1 & 3.25 & 6980 & 2.9854 & 0.295 & 2.37 -- 4.45  & 0.60 & 0.0052 \\
2 & 3.05 & 6552 & 2.9738 & 0.350 & 2.10 -- 4.80  & 0.60 & 0.0066 \\
3 & 2.60 & 6797 & 3.2065 & 0.543 & 2.72 -- 7.60  & 0.91 & 0.0110 \\
4 & 2.30 & 7000 & 3.3994 & 0.767 & 3.98 -- 10.80 & 1.34 & 0.0257 \\
5 & 2.15 & 6604 & 3.3959 & 0.934 & 3.98 -- 11.14 & 1.34 & 0.0263 \\
6 & 2.04 & 6789 & 3.6332 & 1.046 & 4.84 -- 16.24  & 2.07 & 0.0612\\
7 & 1.84 & 6784 & 3.7890 & 1.266 & 5.7 -- 22.5 &  2.72  & 0.1103 \\
\hline
A & 2.21 & 7285 & 3.6488 & 0.812 & 6.8 -- 17.7 & 2.05 & 0.0623\\
B & 2.12 & 7303 & 3.6538 & 0.960 & 7.26 -- 17.3  & 2.10 & 0.0647 \\
C & 2.13 & 7308 & 3.6492 & 0.947 & 7.17 -- 17.11 & 2.07 & 0.0635\\
D & 2.14 & 7305 & 3.6459 & 0.934 & 8.20 -- 16.8 & 2.06 & 0.0626\\
E & 2.04 & 7295 & 3.6460 & 1.076 & 7.6 -- 17.1 & 2.10 & 0.0642\\
F & 1.97 & 7300 & 3.8240 & 0.996 & 8.3 -- 24.0 & 2.80 & 0.1203 \\
G & 1.97 & 7300 & 3.7094 & 1.190 & 6.9 -- 19.5 & 2.34 & 0.0810  \\
H & 1.97 & 7301 & 3.7079 & 1.191 & 6.7 -- 20.83 & 2.34 & 0.0806 \\  
\hline
\end{tabular}
\end{table*}

\begin{figure}
   \centering
   \includegraphics[width=\hsize]{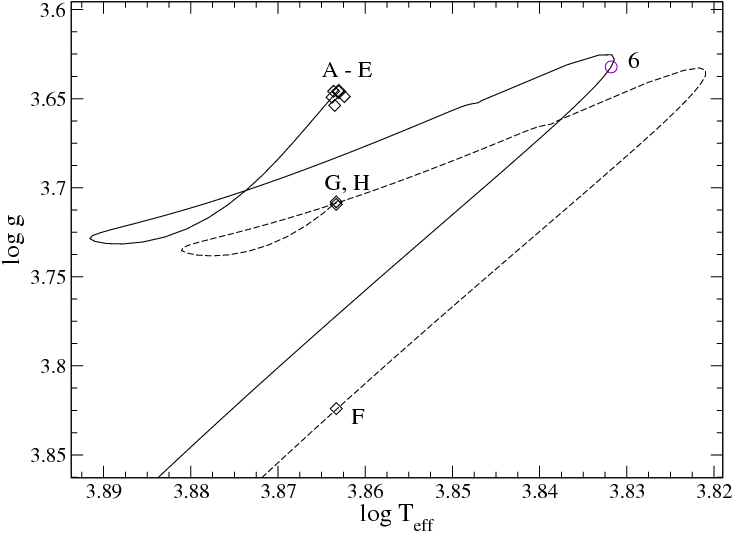}
      \caption{Location of models for component B, in the log T$_{\mathrm{eff}}$ -- $\log g$ plane. The full line corresponds to the evolutionary track of the M = 2.04\,$M_{\sun}$ model, and the dashed line to M = 1.97\,$M_{\sun}$.}
     \label{fig-models-largesep}
\end{figure}

The discrepancies in T$_{\mathrm{eff}}$ and $\log g$ obtained from spectroscopy (Sec.\,\ref{sec-spec}) and orbital solution (Sec.\,\ref{sec-orbit}) led us to place this binary system in a wide range within the HR diagram: T$_{\mathrm{eff}}$ = 6600 - 7300 K and $\log g$ = 3.0 - 3.7. The range of the frequency content [5-15d$^{-1}$ ] (Sec.\,\ref{sec-lightcurve}) and the characteristic spacing (2.0 -- 2.2\,d$^{-1}$) between the dominant modes (assuming that it corresponds to a large separation feature in the low frequency regime) add additional constraints for the pulsating component in the binary system. We aimed at modelling component B to check if one of these observables might tip the scales in favour of the spectroscopic or to the orbital solution. 

We used the evolutionary code CESAM \citep{Morel97, Morel08} and the pulsation code GRACO  \citep{Moya04, Moya08} to calculate non-rotating equilibrium models and non-adiabatic frequencies with l $\in$ [0, 3], respectively. All the models included OPAL opacity tables and Eddington (radiative) atmospheres. The element abundances determined in Sec.\,\ref{sec-abund} have been taken as parameters for the initial chemically homogeneous ZAMS models, which resulted in a metallicity of [Fe/H] = 0.09 dex. The mixing-length formulation \citep{Bohm58} have been included in the computations with the overshooting extension. Following the works by \citet{Daszynska05} and  \citet{Casas06, Casas09} concerning $\delta$ Sct\ pulsators, we used the most consistent values (with observations) for the mixing-length parameter and core overshooting: $\alpha_{\mathrm{MLT}}$ = 0.5 and d$_{ov}$ = 0.2, respectively.

Two set of models covering the spectroscopic (1-5 in Table\,\ref{tab-models}) and the orbital solutions (A-H in  Table\,\ref{tab-models}) were studied but we explored  models, too, using the spectroscopic T$_{\mathrm{eff}}$ and $\log g$ from the orbital solution (6-7 in Table\,\ref{tab-models}). Usually, $\log g$ and T$_{\mathrm{eff}}$ were the fitted parameters, and the mass was considered a free parameter. To cover wider possibilities, in models F, G, and H, the mass and T$_{\mathrm{eff}}$ from the orbital solution were the target parameters, and $\log g$ was a free parameter.

The models cover a large range of evolutionary stages from main sequence (A, F) through the terminal age main sequence (TAMS, 6), the contraction phase (B, C, D), and the H-shell burning phase (E, H) up to the post main sequence phase (PS, 1-5). We calculated the range of excited modes, $\Delta \nu$ (large separation) for low order radial modes, and the mean density ($\rho$) were calculated for all models (Table\,\ref{tab-models}).

Models 1 to 5 show frequency ranges and large separations lower than the observed ones. On the other hand, models F, G, and H possess much higher values for both the range and $\Delta \nu$. Models A, B, C, D, and E  fit the range of excited modes only roughly, but present large separations similar to the observed ones. They still cover the evolutionary phase from MS (A) via contraction phase (B-D) to the H-shell burning phase (E). The most representative model of the star is model 6 with the spectroscopic T$_{\mathrm{eff}}$ and the orbital $\log g$ value. The range of excited modes [4.84-16.24] agree with the observed range and the large separation agrees with the half of the $f_1$ -- $f_4$ spacing. According to this model, the pulsating star is in the TAMS evolutionary stage with 1.046 Gyr age.

Remarkably, even when these latter models that fit the large separation value (A to E and 6) have slightly higher masses than the values in Table\,\ref{tab-rv-orbitparam}, they show very similar mean densities (less than a 2.1\% of difference) to the investigated star's with $\rho$ = 0.0638$\pm$0.0154 g\,cm$^{-3}$, accordingly to values in Table\,\ref{tab-rv-orbitparam}. It has been suggested by several studies, both with \citep{Reese08} or without \citep{Garcia13, Suarez14} the consideration of rotational effects in the frequency computations, that the large separation in the low order regime of the p modes is related in a simple way to the mean density of the star, which is just analogous to the solar-type stars case. For this particular object, the agreement between non-rotating models and observations is due to the moderate rotation.  
Using the orbital solution and the inclination angle, we derived  $\Omega/\Omega_{K}\sim0.135$, where $\Omega_{K}$ is the rotational Keplerian limit ($\Omega_{K} = \sqrt{GM/R^{3}}$).

It is also worth mentioning that the previously commented findings (the large separation and mean density)  match the theoretical formula given in \citet{Suarez14}. We found a deviation of only 6\,\% from the observed values in the parameters calculated with the relation.

\section{Discussion}
\label{sec-discussion}

The comprehensive investigation (CoRoT light curve, spectroscopy, orbital and 
pulsational modelling) of \object{HD\,51844}\ proved to be a challenge in many fields
of astrophysics. Discrepancies appeared between the derived parameters from
different methods, as in the cases of other binary stars \citep[e.g.,][]{Maceroni14, Lehmann13}.

Disentangling of the composite spectra is challenging when the spectra are not evenly spaced in phase like in our case. Finally, the T$_{\mathrm{eff}}$, $\log g$, and the abundances were derived from the composite spectra. The non-LTE effects may have a significant influence
on the balance of \ion{Fe} {I} and \ion{Fe} {II} lines that we used 
for the determination of $\log g$ \citep{Mashonkina11}. 
The difference in T$_{\mathrm{eff}}$ of 500\,K between the spectroscopic determination and the binary 
light curve modelling for the secondary component suggests significant basic uncertainties of the model physics. 

The modelling of the brightening at periastron passage in the light curve was found to be quite sensitive to the type of atmosphere (convective or radiative) of both components.
The choice of either one led to very different basic parameters (gravity darkening and albedo). 

The seismic modelling based on the results of spectroscopy and the orbital solution is also confronted with the problem that the stellar evolution is very rapid in this parameter space. 
The modelling also encounters the problem that according to the abundance analysis both components are Am stars showing a significant Sc deficiency for the pulsating component (secondary). \citet{Maceroni14}  found a similar Sc deficiency for the primary component in the KIC 3858884 system, but, contrary to our case, not for the pulsating component. 

In the case of Am stars, the surface abundances do not represent the mean abundances in the inner layer of the stars \citep[diffusion and settling,][]{Michaud04}.
Moreover, the primordial elemental abundances of the original cloud, from which the system formed, are not reflected by the photospheric abundances. What is more, the capture of the components is unlikely because of the similar abundances.

The best seismic model resulted in a solution with spectroscopic T$_{\mathrm{eff}}$ and with $\log g$ obtained from the orbital solution. Although time dependent convection (TDC) would be needed for a more sophisticated representation of the excitation, the seismic model \#6 perfectly covers the range of the excited modes that we obtained from the CoRoT light curve.

Still, we are faced with the unsolved problem of the non-pulsating 
primary. The line profile variation shows that only the secondary is pulsating.
The direction of the Doppler shift of the four dominant frequencies uniquely proves that they
are excited in the same star, which is consistent with the Doppler shift of the spectral lines of the secondary. 

The wide range of T$_{\mathrm{eff}}$ and $\log g$ of the models cover the parameters of the primary, too, which means we should see pulsation in the primary, but which we could not detect.
It is not a surprising result since about 1/3 of the star shows pulsation in
the classical instability strip.
However, the enhanced abundances of the s-process elements in
both components suggest evolutionary status at or near the post main sequence
phase rather than the above mentioned scenario.

Another argument to why we could not find pulsation in the primary could be low amplitudes or a geometric effect. Since the linear treatment of the non-radial pulsation do not give information on the excited amplitudes, the primary may also pulsate but with a lower amplitude than 
the secondary. The upper amplitude limit may be the uncertainties of the line
profiles. Due to their low amplitudes, we cannot check more frequencies for Doppler shifts; hence, it is not possible to discern directly to which component each frequency belongs.

Different inclination angles of the primary and secondary with similar excited amplitudes are very unlikely, since the stellar rotation and the orbital motion are relics of the same angular momentum of the primordial cloud, out of which the binary was formed. We may suppose that the rotational axes of both components are aligned in the same way to the orbital plane as we supposed in the orbital solution.


As we have discussed in Sec.\,\ref{ssec-freqanal-signal}, 3:2, the 3:5 and 3:4 resonances were found
between pairs of frequencies. Of course, when we have many frequencies, then
numerical relations among the lower amplitude frequencies may appear with
higher probability. However, we have $f_1$:$f_4$=3:2, $f_3$:$f_7$=3:5, and $f_4$:$f_{13}$=3:4
resonances for the higher amplitude modes. The probability that
these ratios appear by chance is very low. We regard their appearance as an up-to-now
unencountered new result, but we did not find any obvious explanation for it. According
to our knowledge, n:9 resonances were found in HR\,8799, a $\gamma$ Doradus star
based on MOST dataset (S\'odor et al., submitted to A\&A).

Three different periodicities exist in a heartbeat binary system: the
orbital period and the rotational periods of the components. We checked the
relation of the frequencies to the orbital period searching for tidally
excited p modes. We found two modes ($f_{22}$ and $f_{36}$) with an integer multiple of the
orbital period. The orbital solution revealed similar but not identical rotational velocities for the components. 

A plausible idea was to check the relations of the frequencies to the rotational velocities. 
Three frequencies ($f_{15}$, $f_{32}$, and $f_{97}$) showed exact integer multiple of the v$_{rot}$ of the secondary component with 28.997422, 38.996355 and 24.000133 values, respectively. Three other frequencies ($f_{31}$, $f_{84}$ and $f_{114}$) revealed exact integer multiple of the v$_{rot}$ of the primary component with 50.002677, 32.000375, and 20.995095 ratios, respectively. A plausible explanation is that not only the orbital period is tidally locked to some p modes, but the rotational velocities of both components are locked to some frequencies. Regarding a $\pm$0.05 range around the exact multiple values, eleven frequencies exhibit near resonance to the orbital frequency, eight to the v$_{rot}$ of the secondary and ten to the v$_{rot}$ of the primary. The frequency $f_{100}$ shows near resonance to both the orbital and the rotational velocity of the secondary in the accepted range.

No frequency was found in resonance with both the orbital and v$_{rot}$ of the primary. In the case of $f_{25}$, near resonance was found in the accepted range to the rotational velocity of both components.
We see only a low probability that so many resonances would appear only by
chance. The frequencies presumably are the eigenmodes of the secondary
component, but they do not seem to have a similar excitation as in a single
pulsating star. We have not found  a similar behaviour reported for any heartbeat star.

The TAMS (terminal age main sequence) evolutionary stage of the secondary
with an age of 1 Gyr, the orbital solution with almost equal rotational velocities, and their relations to the frequencies give information on the evolutionary stage of the binary system.

\citet{Mazeh08} gives a nice review of the tidal effects in binaries, including the eccentricity-period relation, the synchronisation, and circularisation of the systems.
The eccentricity-period relation of HD 51844 agrees with the general distribution of SB binaries \citep[Fig.\,1,][]{Mazeh08}, although it is near to the envelope curves. The medium-range eccentricity (e=0.484) clearly shows that the system is by far not circularised. On the other hand, the equality of the projected rotational velocities suggests that the stellar spins are synchronised with each other.

According to the measured $v_{\rm{rot}}\sin{i}_{\rm{rot}}$ values, the maximum rotation period of the two stars is $\sim4.2$ days, the pseudo-synchronous rotation period would be $\sim12.6$ days while in the frame of the equilibrium tide theory of \citep{Hut81}; therefore, we conclude that the stars certainly exhibit super-synchronous rotation. Furthermore, for such a high ratio of the rotational and orbital angular velocities ($\Omega/\Omega_{orb}\geq8$), the classical theory of equilibrium tides predict eccentricity increase instead of circularisation, as it was pointed out first by \citet{Darwin79}.

In the recent years both the equilibrium \citep{Remus12} and the dynamical \citep[see, e.g.,][and numerous further references therein]{Fuller12, Burkart13} tide theories have been substantially improved. Most of the recent studies concentrated on tidal resonance locking mechanism \citep{Witte99}, which may cause tidally driven rapid synchronisation and circularisation. One of the extraordinary importance of heartbeat stars is that they may offer observational evidence for such tidally-excited oscillations, 
which might indicate tidal resonance locking in progress. The best example is the case of KOI-54 
\citep{Fuller12}. 

The lack of oscillation frequencies in the g-mode region (apart from $f_{98}$ and $f_{111}$) and the appearance of only two p modes, which are multiples of the orbital frequency, reveal the ineffectiveness of the tidal oscillations in the present evolutionary stages of the binary members. The relatively high age of 1\,Gyr of the system also suggests that the resonance locking (which would cause a significant eccentricity decrease on a 0.1\,Gyr timescale) has not been effective for most part of the system's lifetime. However, the relations between the oscillation frequencies and the rotational frequencies of the components could play an important role in reaching a super-synchronous rotation period of the components. The possibly equal super-synchronous 
rotation period of the members of such a highly eccentric, old binary might offer further challenges for
tidal dissipation theories.  

\section{Summary}
\label{sec-conc}

Using the high precision CoRoT Seismo light curve and high-resolution spectroscopy, we discovered that \object{HD\,51844}\ is the first CoRoT heartbeat star. The system consists of two nearly equal mass stars with similar fundamental parameters, showing $\delta$ Sct\ pulsation and integer ratios (resonances) among $\sim$36\% of the pulsation frequencies. Such integer ratios have not been reported before for $\delta$ Sct\ stars. 
For the dominant four frequencies, we could unambiguously show that they originate in the slightly less massive star and could not find proof for the slightly more mass component to pulsate. 
We discovered two potentially tidally excited p modes with an exact integer ratio to the orbital period. Additionally, one less obvious tidally excited g mode was found. We also found an exact locking of frequencies not only to the orbital period but also to the rotational frequencies of both components. No similar behaviour has been reported for any heartbeat star. The rotation of the components is super-synchronised, but the system is far from being circularised.    
The chemical composition of both components fits the $\delta$ Del classification from the literature, where the most pronounced difference is in the Sc abundance ($\sim$ 1 dex less for the pulsating secondary). The equal super-synchronised rotation periods, the high eccentricity, and the age of \object{HD\,51844}\ provide good occasion for further investigation of the tidal dissipation theories. 
   
\begin{acknowledgements}
      MH and MP acknowledge financial support of the ESA PECS project 4000103541/11/NL/KML.
      AGH acknowledges support from the EC Project SPACEINN (FP7-SPACE-2012-312844) and from FCT-MEC (Portugal) through a fellowship ( SFRH/BPD/80619/2011 ). TB would like to thank City of Szombathely for support under Agreement No.S-11-1027.
      MR and EP acknowledge financial support from the FP7 project {\it SPACEINN: Exploitation of Space Data for Innovative Helio-and Asteroseismology}. 
      WWW acknowledges financial support form the Austrian Science Fonds (FWF P 22691-N16).
      This research has made use of the SIMBAD database, operated at CDS, Strasbourg, France.
      This research has made use of NASA's Astrophysics Data System Bibliographic Services.
\end{acknowledgements}

\bibliographystyle{bibtex/aa}
\bibliography{hd51844-references.bib}

\listofobjects
\onecolumn
\Online

\begin{longtable}{lrcrrrrrlrcrr}
\caption{The 115 frequencies, amplitudes and phases obtained above S/N = 4.0.}\label{tab-online}\\
\hline
\hline
No. & \multicolumn{1}{c}{Frequency} & \multicolumn{1}{c}{Ampl.} & \multicolumn{1}{c}{Phase} &
\multicolumn{1}{c}{S/N} & & & &\multicolumn{1}{c}{No.} & \multicolumn{1}{c}{Frequency} & \multicolumn{1}{c}{Ampl.}
& \multicolumn{1}{c}{Phase} & \multicolumn{1}{c}{S/N}  \\
& \multicolumn{1}{c}{(d$^{-1}$)} & \multicolumn{1}{c}{(mmag)} & \multicolumn{1}{c}{(rad)} &  & & & & &
\multicolumn{1}{c}{(d$^{-1}$)} &  \multicolumn{1}{c}{(mmag)} &
\multicolumn{1}{c}{(rad)} &  \\
\hline
1 & 12.21284(4) & \multicolumn{1}{r}{2.26(2)} & 0.1683(8) & 419.99 & & & & 59 & 5.523(6) & 0.04(8) & 0.46(8) & 11.49 \\
2 & 7.05419(8) &   2.14(2) & 0.2278(9) & 431.56 & & & & 60 & 11.942(9) & 0.05(1) & 0.62(6) & 9.51 \\
3 & 6.94259(6) & 1.83(2) & 0.5071(6) & 371.62  & & & & 61 & 9.283(6) & 0.04(8) & 0.06(9) & 8.73 \\
4 & 8.14077(3) & 1.47(4) & 0.150(7) & 282.77  & & & & 62 & 10.971(0) & 0.04(5) & 0.01(9) & 7.94 \\
5 & 6.75568(5) & 0.61(5) & 0.797(7) & 125.20 & & & & 63 & 12.050(0) & 0.04(3) & 0.35(9) & 7.95 \\
6 & 11.95900(1) & 0.58(4) & 0.388(8) & 108.89  & & & & 64 & 9.072(3) & 0.04(3) & 0.76(6) & 7.96 \\
7 & 11.56613(9) & 0.53(7) & 0.512(1) & 95.27  & & & & 65 & 9.470(0) & 0.04(1) & 0.44(8) & 7.58 \\
8 & 10.3695(3) & 0.37(1) & 0.476(9) & 66.53 & & & & 66 & 7.084(2) & 0.04(2) & 0.53(7) & 8.43 \\
9 & 11.3373(2) & 0.36(5) & 0.440(8) & 65.74 & & & & 67 & 11.316(5) & 0.04(3) & 0.57(3) & 7.66 \\
10 & 12.0898(2) & 0.35(7) & 0.554(2) & 65.86 & & & & 68 & 6.769(9) & 0.04(3) & 0.82(1) & 8.82 \\
11 & 10.3528(3) & 0.32(3) & 0.292(2) & 57.71 & & & & 69 & 11.738(8) & 0.03(9) & 0.89(4) & 6.88 \\
12 & 10.7233(6) & 0.30(6) & 0.970(0) & 55.41 & & & & 70 & 10.260(6) & 0.03(9) & 0.41(2) & 7.01 \\
13 & 10.8483(5) & 0.27(7) & 0.957(2) & 49.87  & & & & 71 & 9.776(1) & 0.03(9) & 0.01(0) & 7.13 \\
14 & 10.6819(2) & 0.26(1) & 0.687(6) & 47.53 & & & & 72 & 5.087(4) & 0.03(7) & 0.44(5) & 9.26 \\
15 & 7.4543(5) & 0.23(7) & 0.359(2) & 46.27 & & & & 73 & 12.466(8) & 0.03(8) & 0.88(9) & 7.14 \\
16 & 11.7759(5) & 0.22(9) & 0.230(9) & 40.85 & & & & 74 & 6.972(6) & 0.03(8) & 0.85(0) & 7.60 \\
17 & 12.5201(9) & 0.19(1) & 0.506(8) & 36.23  & & & & 75 & 8.169(6) & 0.03(7) & 0.18(6) & 7.00 \\
18 & 11.4296(6) & 0.18(3) & 0.132(8) & 32.48  & & & & 76 &  7.913(9) & 0.03(7) & 0.72(9) & 7.08 \\
19 & 6.5868(9) & 0.17(2) & 0.360(1) & 35.92 & & & & 77 & 6.547(4) & 0.03(2) & 0.66(7) & 6.72 \\
20 & 12.7545(3) & 0.17(2) & 0.064(6)& 33.45 & & & & 78 & 7.275(9) & 0.03(2) & 0.14(6) & 6.43 \\
21 & 11.8311(5) & 0.16(2) & 0.07(6) & 29.07 & & & & 79 & 10.210(6) & 0.03(2) & 0.44(9) & 5.62 \\
22 & 9.9707(4) & 0.16(3) & 0.86(1) & 28.81 & & & & 80 & 8.690(3) & 0.03(0) & 0.86(5) & 5.84 \\
23 & 7.2190(1) & 0.15(5) & 0.94(3) & 31.35 & & & & 81 & 10.002(9) & 0.03(3) & 0.18(3) & 5.74 \\
24 & 9.7186(2) & 0.14(8) & 0.02(8) & 27.23 & & & & 82 & 9.802(9) & 0.03(2) & 0.69(8) & 5.92 \\
25 & 11.2958(9) & 0.14(3) & 0.90(8) & 25.63 & & & & 83 & 11.656(7) & 0.03(0) & 0.20(7) & 5.34 \\
26 & 14.8846(6) & 0.13(9) & 0.70(6) & 24.64 & & & & 84 & 11.458(6) & 0.03(2) & 0.65(9) & 5.59 \\
27 & 9.3199(5) & 0.12(9) & 0.85(0) & 23.70 & & & & 85 & 8.624(0) & 0.03(0) & 0.09(5) & 5.84 \\
28 & 8.8682(8)  & 0.13(8) & 0.07(9) & 26.47  & & & & 86 & 11.505(2) & 0.03(0) & 0.94(6) & 5.33 \\
29 & 8.8818(8) & 0.12(2) & 0.17(3) & 23.42 & & & & 87 & 9.917(9) & 0.02(9) & 0.11(1) & 5.23 \\
30 & 12.1572(4) & 0.11(2)  & 0.25(6) & 20.65 & & & & 88 & 8.545(7) & 0.02(8) & 0.12(4) & 5.45 \\
31 & 12.2856(7) & 0.10(9)  & 0.09(6) & 20.20 & & & & 89 & 7.862(5) & 0.02(7) & 0.74(0) & 5.17 \\
32 & 10.0247(7) & 0.10(1)  & 0.25(3) & 17.91 & & & & 90 & 6.518(7) & 0.02(7) & 0.78(1) & 5.90 \\
33 & 8.3888(4) & 0.09(9) & 0.45(5) & 19.06 & & & & 91 & 10.080(2) & 0.02(7) & 0.31(3) & 4.83 \\
34 & 12.4223(7) & 0.10(0)  & 0.69(9) & 18.72 & & & & 92 & 10.523(1) & 0.02(7) & 0.06(6) & 4.95 \\
35 & 7.3205(4) & 0.09(2)  & 0.26(9) & 18.32 & & & & 93 & 10.837(4) & 0.02(7) & 0.21(3) & 4.87 \\
36 & 9.5227(3) & 0.08(6)  & 0.73(0) & 15.67 & & & & 94 & 9.460(6) & 0.02(7) & 0.09(5) & 4.88 \\
37 & 10.6712(7) & 0.08(6) & 0.72(6) & 15.71 & & & & 95 & 8.915(1) & 0.02(9) & 0.68(6) & 5.42 \\
38 & 10.5005(0) & 0.08(3)  & 0.06(2) & 14.96 & & & & 96 & 5.677(6) & 0.02(6) & 0.27(0) & 6.21 \\
39 & 7.9010(9) & 0.08(4)  & 0.82(3) & 16.07 & & & & 97 & 6.169(7) & 0.02(5) & 0.04(0) & 5.68 \\
40 & 11.0848(4) & 0.07(6)  & 0.28(5) & 13.22 & & & & 98 & 2.873(7) & 0.02(5) & 0.68(4) & 4.94 \\
41 & 11.2855(7) & 0.08(0)  & 0.60(2) & 14.28 & & & & 99 & 11.606(1) & 0.02(5) & 0.27(9) & 4.50 \\
42 & 12.2430(9) & 0.07(2)  & 0.58(3) & 13.33 & & & & 100 & 7.700(6) & 0.02(4) & 0.87(8) & 4.60 \\ 
43 & 7.5712(2) & 0.07(1)  & 0.60(1) & 13.86 & & & & 101 & 6.068(1) & 0.02(3) & 0.14(7) & 5.24 \\
44 & 12.1830(0) & 0.07(1)  & 0.96(0) & 13.09 & & & & 102 & 6.300(3) & 0.02(4) & 0.45(4) & 5.36 \\
45 & 7.6520(5) & 0.06(7)  & 0.83(3) & 12.97 & & & & 103 & 10.280(2) & 0.02(4) & 0.00(6) & 4.24 \\
46 & 6.7898(5) & 0.06(3)  & 0.04(9) & 12.85 & & & & 104 & 9.857(1) & 0.02(2) & 0.49(4) & 4.10 \\
47 & 8.2485(4) & 0.06(7)  & 0.18(1) & 12.75 & & & & 105 & 10.554(7) & 0.02(3) & 0.71(8) & 4.10 \\
48 & 9.0977(1) & 0.06(3)  & 0.90(0) & 11.82 & & & & 106 & 7.752(5) & 0.02(3) & 0.72(6) & 4.36 \\
49 & 12.7807(9) & 0.06(0) & 0.55(3) & 11.63 & & & & 107 & 12.498(7) & 0.02(5) & 0.94(6) & 4.73 \\
50 & 12.3208(9) & 0.05(9) & 0.53(2) & 11.00 & & & & 108 & 11.813(0) & 0.02(2) & 0.75(1) & 3.98 \\
51 & 10.8197(9) & 0.05(5) & 0.12(1) & 9.83 & & & & 109 & 5.695(7) & 0.02(0) & 0.16(4) & 4.83 \\
52 & 6.9123(6) & 0.05(7) & 0.59(7) & 11.52 & & & & 110 & 12.477(0) & 0.02(1) & 0.16(1) & 4.00 \\ 
53 & 7.0245(0) & 0.05(3) & 0.53(5) & 10.66 & & & & 111 & 3.253(4) & 0.01(9) & 0.32(5) & 4.09  \\
54 & 8.1106(2) & 0.05(2) & 0.96(6) & 10.02 & & & & 112 & 5.557(0) & 0.01(9) & 0.07(3) & 4.75 \\
55 & 8.0741(5) & 0.05(1) & 0.02(1) & 9.85 & & & & 113 & 5.793(5) & 0.01(9) & 0.13(8) & 4.53 \\
56 & 10.306(5) & 0.04(9) & 0.42(7) & 8.77 & & & & 114 & 5.158(5) & 0.01(7) & 0.17(5) & 4.29 \\
57 & 10.867(5) & 0.04(9) & 0.34(1) & 8.89 & & & & 115 & 4.988(5) & 0.01(7) & 0.18(6) & 4.12 \\
58 & 10.885(2) & 0.04(9) & 0.73(4) & 8.73 \\

\end{longtable}
 
\end{document}